\documentclass[smallextended]{svjour3}       % onecolumn (second format)
\smartqed  % flush right qed marks, e.g. at end of proof
\usepackage{graphicx}
\usepackage[pdfborder={0 0 0}]{hyperref}
\usepackage{placeins}
\usepackage{amsmath}
\usepackage{amsfonts}
\usepackage{acronym}
\usepackage{natbib}
\usepackage{color}
\usepackage{authblk}
\usepackage{blindtext}
\usepackage{url}
\usepackage{hyperref}
\journalname{accepted version}

\begin{document}

\title{Mobility Functional Areas and COVID-19 Spread}

\titlerunning{Mobility Functional and COVID-19}        % if too long for running head

\author{Stefano Maria Iacus \and
        Carlos Santamaria \and
        Francesco Sermi \and
        Spyridon Spyratos \and
        Dario Tarchi \and
        Michele Vespe
}

\institute{S.M. Iacus (\emph{Corresponding author}), C. Santamaria, F. Sermi, S. Spyratos, D. Tarchi, M. Vespe  \at
              European Commission\\
              Joint Research Centre\\
              Via Enrico Fermi 2749, 21027 Ispra (VA), Italy\\
              \email{stefano.iacus@ec.europa.eu}           
}

\authorrunning{Iacus, Santamaria, Sermi, Spyratos, Tarchi, Vespe} % if too long for running head

\date{Received: xxx / Accepted: xxx}

\maketitle

 \begin{abstract} 
 This work introduces a new concept of functional areas called Mobility Functional Areas (MFAs), i.e., the geographic zones highly interconnected according to the analysis of mobile positioning data.
 The MFAs do not coincide necessarily with administrative borders as they are built observing natural human mobility and, therefore, they can be used to inform, in a bottom-up approach, local transportation, spatial planning, health and economic policies. 
%Compared to \textit{ad hoc} case studies and  survey analyses, this big data technique can be used to extract MFAs at higher time frequency and spatial resolution. Moreover, looking at past data it is possible to estimate the impact of policies or exogenous shocks to mobility like in the case of the COVID-19 pandemic.
After presenting the  methodology behind the MFAs, this study focuses on the link between the COVID-19 pandemic and the MFAs in Austria. It emerges that the MFAs registered an average number of infections statistically larger than the areas in the rest of the  country, suggesting the usefulness of the MFAs in the context of targeted re-escalation policy responses to this health crisis.  The MFAs dataset is openly available to other scholars for further analyses.

\end{abstract}

%\section*{Highlights}
%\begin{itemize}
%\item Human mobility naturally shapes MFAs in time and space;
%\item lockdown measures have shown an overall ``shrinking" effect on the MFAs across Europe;
%\item MFAs are persistent in time with intra-weekly recurrent patterns
%\item MFAs can be used to inform targeted mobility containment measures in case of new outbreaks, providing a balance between epidemiological and socio-economic impact.
%\end{itemize}

%\newpage

\section{Introduction}

In April 2020, the European Commission (EC) asked European Mobile Network Operators (MNOs) to share fully anonymised and aggregated mobility data in order to support the fight against COVID-19 \citep{EURec, Roadmap} with data driven evidence.

The value of mobile positioning \textit{personal data} to describe human mobility has been already explored in literature \citep{csaji2013exploring} and its potential in epidemiology  demonstrated  \citep{wesolowski2012quantifying,jia2020population,Wu,kraemer2020effect}.
The new initiative between the Commission and the European MNOs relies on the effectiveness of using \textit{fully anonymised and aggregated mobile positioning data} in compliance with `Guidelines on the use of location data and contact tracing tools in the context of the COVID-19 outbreak' by the European Data Protection Board \citep{edpb2020}.

This work makes use of mobile data to introduce a new concept of `functional areas' called \textit{Mobility Functional Areas} or MFAs. 
The concept of functional areas  has a long tradition in settlement geography, urban planning and policy making \citep{Ball80,VanderLaan98,Casado2000,OECD2002,Andersen02,Estat2016,dMFA,OECD2019}. The idea behind functional areas is the identification of a network of aggregated inbound and outbound movements across spatial structures for a given time scale (for example, daily, intra-weekly, seasonally, etc) according to the scopes of their use.

Thanks to the above mentioned unprecedented collaboration between the European Commission and European Mobile Network Operators during the COVID-19 ongoing pandemic, it has been possible to define the MFAs for 15 European countries  (14 member states of the European Union: Austria, Belgium, Bulgaria, Czechia, Denmark, Estonia, Spain, Finland, France, Greece, Croatia, Italy, Sweden, Slovenia, plus Norway),  although in this study we will focus on the case of Austria. 

Despite the fact that  mobility data alone cannot predict future needs \citep{Bwambale20}, they can show already compelling citizens needs, like  transportation or health care facility allocation needs. Moreover, thanks to the capability of collecting mobile data at very high time frequency and space granularity and backward in time, the time evolution of the MFA can indeed show changes or ongoing trends or help to design policies or measure the effectiveness of policies.

This work is organised as follows. Section~\ref{sec:fa}  is a brief summary of the different approach to the definition of functional areas and explains the importance of this concept in the field of policy making. Section~\ref{sec:data} describes the essential characteristics of the mobile data used in the analysis. Section~\ref{sec:MFA} explains in details the method of construction of the MFAs starting from mobility data. As the MFAs are constructed on a daily basis  and slightly change daily due to, e.g., intra-weekly human mobility patterns, Section~\ref{sec:persistent}  describes a denoising method to extract the subset of areas which are consistently connected  stable  through time. In this respect, the  pre-lockdown (or persistent) and lockdown  MFAs generated by COVID-19 mobility restriction measures are also discussed.  Section~\ref{sec:Vienna} presents several potential applications of the concept of Mobility Functional Areas, including transport system, socio-economic analysis, public health, urban planning and environmental risk analysis.
To show the effectiveness of the MFAs concept to policy making,  Section \ref{sec:austria}  focuses on a public health application. It analyses the impact of the MFAs on the spread of the coronavirus during the first wave of the pandemic  in  Austria  showing that the number of COVID-19 cases increases more within the MFAs than outside. Section \ref{sec:end} discusses limits and further potential usage of MFAs. The MFAs dataset is openly available to other scholars.

\section{Functional areas and policy making}\label{sec:fa}
As mentioned, the concept of functional areas has a long tradition in settlement geography, urban planning and policy making. These areas have and can be defined in several ways and for different purposes e.g. transportation planning, urban governance and regional development among others. Our proposed concept of highly interconnected geographic zones, the MFAs, are purely data driven were originally developed to fight the COVID-19 pandemic in Europe. Since MFAs are data-driven products generated by natural human mobility, they can be used for several other applications (see Section~\ref{sec:Vienna}). A few pre-existing variants of the same concept are the following:
\begin{itemize}
	\item \textit{`commuting regions'}: the identification of relatively closed regions of daily moves of residing population based on commuting data from censuses \citep{Casado2000,VanderLaan98}. The aim of the commuting regions is to define geographical areas within which the majority of the work to home travels occur. The definition of commuting regions is particularly relevant for housing and transportation planning purposes.
	\item \textit{`functional regions'}: a tool used to target areas of specific national and European policies \citep{OECD2002}. There are several natural areas of application of functional regions including employment and transportation policies, environmentally sustainable spatial forms, reforms of administrative regions, strategic level of urban and regional planning and a wide range of geographical analyses (migration, regionalisation, settlement system hierarchisation) \citep{Andersen02,Ball80,Casado2000,VanderLaan98}.
	\item \textit{`functional urban areas'}: cities with their commuting zone \citep{Estat2016,OECD2019}. They are generally identified by a densely inhabited city, together with a less densely populated commuting zone whose labour market is highly integrated with that of the city. They reflect the functional and economic extent of cities and not just their administrative extent. The Functional Urban areas are thus the appropriate spatial level for a more comprehensive urban governance and planning.
	\item \textit{`overlapping functional regions'} by \cite{Killer2010}. The overlapping functional regions reflect the complexity of commuting flows. One area can be linked to several others and thus could belong to multiple functional areas.
\end{itemize}
 More in general, there exists a long tradition in defining zones related to human mobility. In their extensive review \cite{PattersonFarber2015} summarize more than 61 different ways of defining Potential Path Areas (PPAs) and Activity Spaces (ASs). The PPAs refer to 'the spatial extent of where individuals can participate in activities subject to constraints' (e.g., time, modal availability, etc);
ASs refer to locations ‘with which individuals have direct contact as the result of day-to-day activities’. Our MFAs concept, by design, belongs to the ASs strand of research, though MFAs can be potentially used also in the context of PPAs using a probabilistic setup, but this goes beyond the scope of the present study. A further strand of research is that of Transportation Analysis Zones (TAZs), which is generally (though not exclusively) aimed at the optimization of the transport network at suburban level using very high resolution data and geographical/administrative constraints. Unfortunately, the types of aggregated data of this project do not allow this specific application, though high resolution MNO data (XDR or CDR data) may fit the purpose using the same technique applied in this study.
As highlighted in \cite{PattersonFarber2015}, most of these methods have in common some types of calculation like, e.g., ellipse methods (Standard Deviation Ellipses, Travel Probability Fields, etc.),  Kernel methods (Kernel Density
estimation), network analysis, location methods (specific locations at which activities are performed, for example, shopping malls or trip destinations, Points of interests), Minimum Convex-hull Polygons (MCP), etc. Our MFAs rely on the network analysis approach.

 Back to the concept of functional areas, the most common data sources for the above-mentioned studies are by far the population censuses and \textit{ad hoc} pilot surveys.
In these approaches travel behaviour is captured collecting costly household travel survey data, which are usually very small in terms of sample size,  rarely updated and sometimes  prone to reporting errors \citep{Bwambale20}.

Mobile data  have been used in the past in some pilot-studies like in Estonia \citep{Novak2013} or in some regions of Italy in \citet{dMFA}.  An attempt to mix survey, census  and mobile data has been proposed as a proof-of-concept by \citet{Bwambale20} for the case of Bangladesh.

 Very recently Facebook released the concept of Commuting Zones\footnote{Details are available here: \url{https://dataforgood.fb.com/tools/commutingzones/}.} (CZs)  within their data For Good project. This CZs  make use of data coming from the portion of Facebook's user base who opt-in to share their location in the mobile app. The construction of the CZs is also based on a network analysis approach.

The approach used to build the MFAs presented in the next section is special in several ways. Indeed, the approach has been designed to be robust to the diversified data sources available for the 15 countries and also in that it tries to take into account persistent patterns of mobility rather than focusing on a single period as we know that mobility patterns may change due to seasonalities or because there is a strong intra-weekly periodicity\footnote{For example, Facebook's CZs are built on the aggregated movements of three months period.}. This was possible thanks to the long time series made available through the unprecedented agreement.

As for the other functional areas that make use of mobility data, the MFAs are particularly interesting as the underlying data can be collected at high frequency compared to any survey data, and also covering the whole territory of a country rather than focusing on special cities or routes.
Remark also that, since mobile phone services unique subscribers\footnote{All  mobile services subscribers, including IoT, are about 86\% of the population, 76\% of which real smartphone users.} represent about 65\% 
of the population across Europe \citep{GSMA2020}, mobile data can reliably be used to capture the aggregate mobility patterns.

In a policy making framework especially related, but not limited, to the COVID-19 pandemic, the insights resulting from the identification of these new mobility areas may help governments and authorities at various levels:
\begin{itemize}
	\item[a)] to control virus spread by limiting non-essential movements outwards an MFA with a high virus infection rate compared to its neighbour MFAs especially in the initial phase of a  virus outbreak;
	\item[b)] to apply targeted and local physical distancing policies in different MFAs, according to their specific epidemiological situation, thus minimizing virus spread and at the same time limiting the economic and social impact of such measures by avoiding not adequate national and universal physical distancing policies;
\end{itemize}

In the absence of any other information, most of the governments are forced to use administrative areas, such as regions, provinces and municipalities to impose physical distancing measures and mobility restrictions. Nevertheless, administrative boundaries are static and do not reflect actual mobility. On the other hand, both the potential spreading of the virus and the territorial economy strongly depend on local mobility \citep{TR-Connectivity}.

Although not all of these aspects can be taken into account in this work, the hypothesis is that the implementation of different physical distancing strategies (such as school closures or other human mobility limitations) based on MFA instead of administrative borders might lead to a better balance between the expected positive effect on public health and the negative socio-economic fallout for the country.  To this aim, Section~\ref{sec:austria} will present a the case study of Austria focusing on the impact of MFAs on the spread of the coronavirus.

\section{Mobile positioning data}\label{sec:data}
The agreement between the European Commission and the MNOs defines the basic characteristics of fully anonymised and aggregate data to be shared with the Commission's Joint Research Centre (JRC\footnote{The Joint Research Centre is the European Commission's science and knowledge service. The JRC employs scientists to carry out research in order to provide independent scientific advice and support to EU policy.}). 
%The JRC processes the heterogeneous sets of data from the MNOs and creates a set of mobility indicators and maps at a level suitable to study mobility comparatively across countries; this level is referred to as `common denominator'.\\
This section briefly describes the original mobile positioning data from the MNOs; the following section introduces the mobility indicator derived by JRC and used in this research.

Data from MNOs are provided to JRC in the form of Origin-Destination-Matrices (ODMs) \citep{ODMs2019, ODMs2020}. Each cell  $[i,j]$ of the ODM shows the overall number of \textit{`movements'} (also referred to as `trips' or `visits') that have been recorded from the origin geographical reference area $i$ to the destination geographical reference area $j$ over the reference period.
In general, an ODM is structured as a table showing:
\begin{itemize}
	\item reference period (date and, if available, time);
	\item area of origin;
	\item area of destination;
	\item count of movements.
\end{itemize}
Despite the fact that the ODMs provided by different MNOs have similar structure, they are often very heterogeneous. Their differences can be due to the methodology applied to count the movements, to the spatial granularity or to the time coverage. Nevertheless, each ODM is consistent over time and relative changes are possible to be estimated. 
%This allows defining common indicators (such as `mobility indicators' \cite{TR-Indicators}, `connectivity matrices' \cite{TR-Connectivity} and `mobility functional areas' that can be used, with all their caveats, by JRC in the framework of this joint initiative.

%Although the ODM contains only anonymised and aggregate data, in compliance with the EDPB guidelines \cite{edpb2020}, upon the reception of each ODM, the JRC carries out a \textit{`Reasonability Test'}. Both the reasonability test and the processing of the ODM to derive mobility indicators take place within the JRC's \textit{Secure Platform for Epidemiological Analysis and Research} (SPEAR).

\section{Mobility functional areas}\label{sec:MFA}
The construction of the MFAs starts from the ODM at the highest spatial granularity available.  This project handled data with different resolution: municipalities, postal code areas, census areas, and different type of grid-level areas. Table~\ref{tab:MFAdata} in the Appendix reports in details the characteristics of the different sets of data used for to calculate the MFAs.
The methodology to derive the MFAs is the same  for all the 15 countries for which the data are available, though in this study we will focus on Austria only. As it will be explained below, the proposed method is simple with the aim of being robust to the different data specification. Indeed, even for those countries for which the data are made at disposal by more than one MNO, the final MFAs are quite similar. %In this study we focus on the first wave of the pandemic and thus the data considered end in June 2020.
The analysis in this study considers only the mobility from Monday to Friday (with the exclusion of festivities) and this is because weekdays  are more relevant for, e.g., transportation system and socio-economic analysis, though in practice the construction of the MFAs can be done for holidays and weekend days with exactly the same steps.
The construction of the MFAs starts from origin-destination matrix. 
Let $ODM_{d,i,j}$ an element of the ODM matrix for date $d$ representing the number of movements from  area $i$ to  area $j$, $i,j=1, \ldots, n$ 
$$ODM^*_{d,i,j} = \frac{ODM_{d,i,j}}{\sum\limits_{j=1}^n ODM_{d,i,j}}, \quad i,j, = 1,\ldots, n.$$
where $n$ is the total number of rows and columns of the ODM (which is a $n\times n$ matrix) and let $ODM^*_{d,i,j}$ the corresponding element of the ODM normalised by row.\\
Now we transform the  $ODM^*_d$ matrix into a 0/1 proximity matrix $P_d$ as follows
$$P_{d,i,j} = \begin{cases} 
1,& \quad ODM^*_{d,i,j} > \textrm{threshold},\\
0,& \quad \textrm{otherwise},
\end{cases}
$$
where the threshold has been set to 15\% and  30\% to derive two types of MFAs, that we will call \textit{regular} or \textit{strict} respectively.
The threshold of 15\% was suggested by  \cite{Novak2013} but also used in the functional urban areas concept developed by  \cite{Estat2016} and \cite{OECD2019}.  This threshold is a tuning parameter that characterizes the final shapes of the the MFAs. We tested different thresholds above and below 15\% including a uniform distribution threshold (i.e., the threshold is set dynamically as the value corresponding to the frequency of a uniform distribution of movements from a given origin to all destinations from this origin that we see in a row of the ODM). Empirically, we can confirm that the 15\%  seems to be the most effective in isolating stable MFAs for all the countries analysed. It is important to remark that our MFAs are not guided by a theoretical approach as they are completely data driven. Other approaches could have imposed an objective function, like, a minimal number of movements, the maximal length, etc and run an optimization algorithm on top of all the steps presented below to find the \textit{optimal} threshold. So in this study 15\% and 30\% are informed \textit{a priori} choices.
 But the main argument on fixing a threshold is that we want to isolate directional movements rather ellipse-like areas. Indeed, the higher the threshold the more the MFAs is concentrated along specific directions.
One further remark is that by using relative number of movements, we do not discriminate MFAs based on their absolute volume of movements/population. This is an information that can be checked ex-post.

As the ODM matrix is not symmetric, so is the proximity matrix, which is transformed into an adjacency matrix $A$ through the following expression:
$$
A_d= \frac12 \cdot \left(P_d + P_d^T\right)
$$
so that each element of $A_d$ can take only three values:
\begin{itemize}
	\item $A_{d,i,j}=0$ if there are no movements from $i$ to $j$ and viceversa (i.e. the two areas are not connected);
	\item $A_{d,i,j}=0.5$ if there are movements only in one direction, either $i$ to $j$ or $j$ to $i$.
	\item $A_{d,i,j}=1$ if there are movements in both directions, from $i$ to $j$ and from $j$ to $i$;
\end{itemize}
From the adjacency matrix we construct a directed\footnote{An undirected graph could be used as well, but we use a directed graph in view of the community detection algorithm used later on.} graph where the vertex represent the  areas  $i=1, \ldots, n$ and the edges are weighted according to the matrix $A$. The MFAs are calculated using a community detection technique called \textit{walktrap} algorithm \citep{PonsLatapy2006}, which finds communities through a series of short random walks.  This approach is different from the \textit{intramax} algorithm used in \citep{Killer2010,Novak2013}. The idea is that these random walks tend to stay within the same community. The goal of the walktrap algorithm is indeed to identify the partition of a graph that maximises its modularity.  The modularity of a graph is the index designed to measure the strength of division of a network into modules (also called groups, clusters or communities). This is why the walktrap algorithm is well suited for finding clusters of fully interconnected cells where most of the movements are internal.\\
All the communities with only one member, i.e. those without inbound and outbound movements over  the 15\% (or 30\%) threshold,  are collapsed into a single big fictitious area representing the territory that either cannot be identified as a pure MFA, or it is just a collection of atomic (mobility-wise)  areas.
As this algorithm is data-driven, the number of MFAs is not prescribed in advance but is the outcome of the algorithm itself. For example, for Austria we have more than 200 MFAs per day, some of them are very small, others are quite large. In all events, their shapes depend only on the mobility itself.

\subsection{Mobility patterns and mobility functional areas}
It is well known and expected that mobility changes between weekdays, weekends and holidays, but there might be also an internal variability within the working week as well as across weeks (e.g., not all Mondays are \textit{exactly} the same in terms of mobility).  Therefore, the MFAs might be slightly different from day to day. In order to denoise the MFAs we will first measure how much the MFAs are stable and consistent through time through a similarity index.

\begin{figure}[!htb] 
    \centering
    \includegraphics[width=1\textwidth]{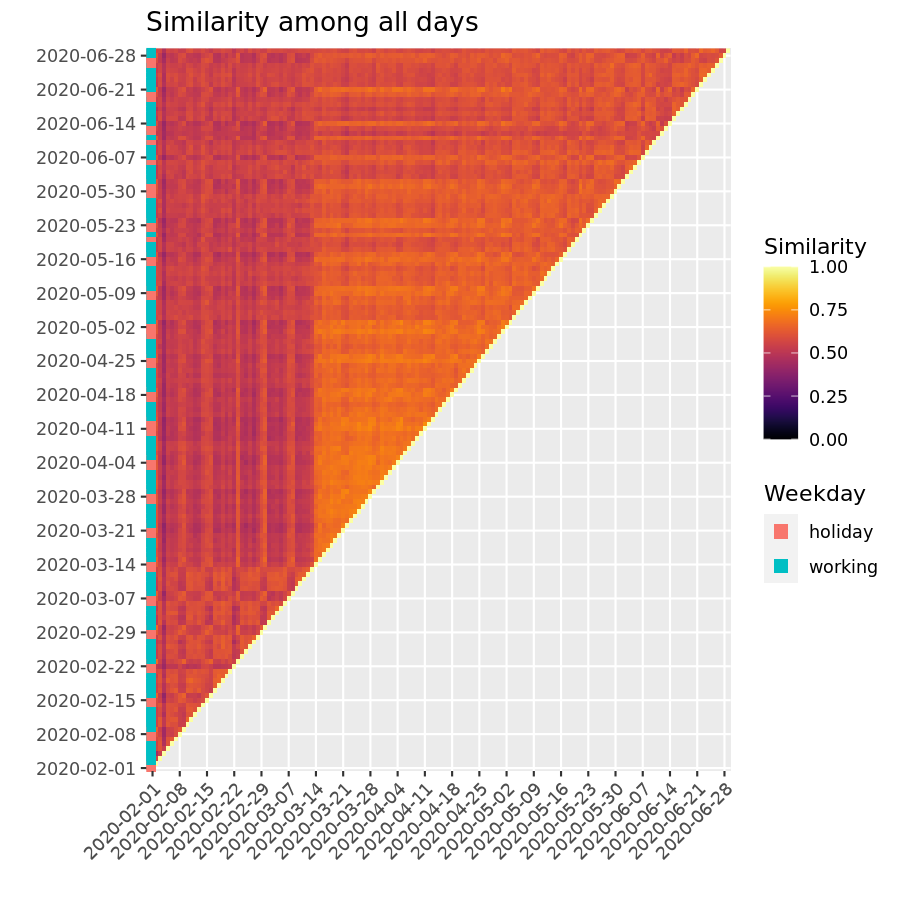}
    \caption{Similarity index matrix of all MFAs in Austria. Period: 1 February 2020 - 29 June 2020. It is clear that the  persistent (pre-lockdown) MFAs are different from the lockdown MFAs and that slowly, the mobility is going back to normality by the end of June 2020.  Lockdown was enforced in Tyrol on 13 March 2020 and nationwide on the 16th of March 2020.}
    \label{fig:allDays}
\end{figure}

\begin{figure}[!htb] 
    \centering
    \includegraphics[width=1.1\textwidth]{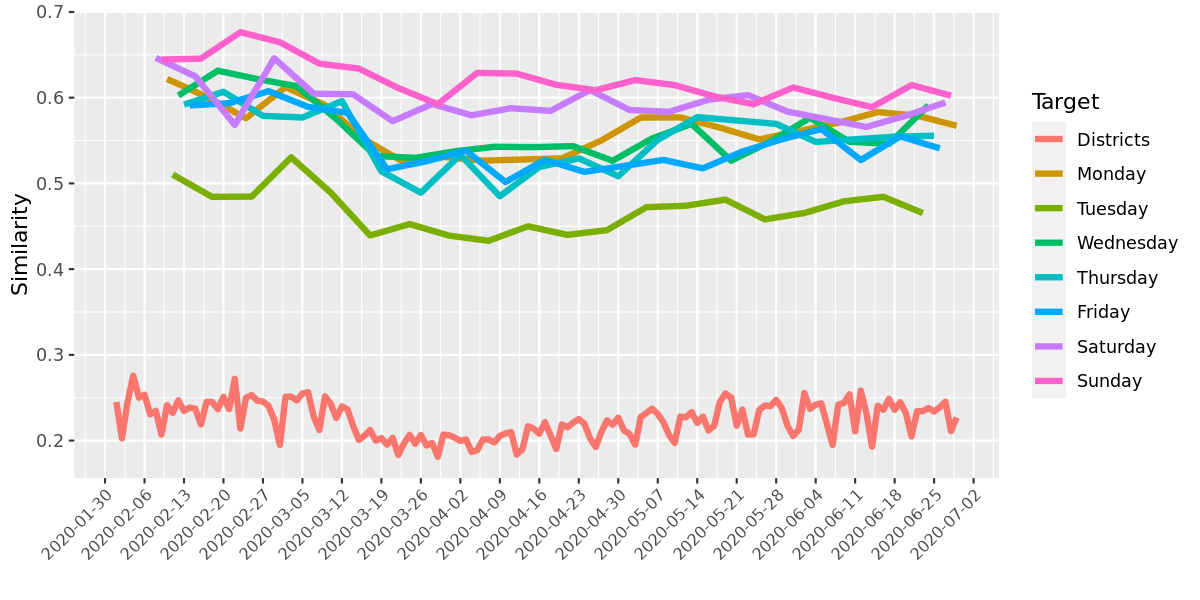}
    \caption{Intra weekly similarity (Mondays versus first Mondays, etc) of daily MFAs and with respect to the Austrian districts (daily MFAs with respect to administrative borders). Lockdown was enforced in Tyrol on 13 March 2020 and nationwide on the 16th of March 2020.}
    \label{fig:similarity}
\end{figure}

We make use of the following similarity index \citep{Gravilov2000} between two sets of groups of labelled $G=\{G_1, \ldots, G_K\}$ and $G' = \{G_1', \ldots, G_{K'}'\}$, where $K$ and $K'$ are not necessarily equal. For example, when two cluster algorithms are applied to the same set of $n$ observations, it might happen that algorithm 1 produces $K$ groups and algorithm 2 produces $K'$ groups.  In this cluster analysis notation, each $G_i$ (or $G_j'$) is a subset of indexes corresponding to observations that fall in group $i$ (respectively $j$). In our case, $G_i$ is the subset of areas id's corresponding to the $i$-th MFA. The similarity index is defined as
\begin{equation}
\textrm{Sim}(G,G') = \frac{1}{K} \sum_{i=1}^k \left\{ \max\limits_{j=1, \ldots,K'} \textrm{sim}(G_i,G_j')\right\}
\label{eq:sim}
\end{equation}
where
$$
\textrm{sim}(G_i,G_j') = 2 \frac{|G_i \cap G_j'|}{|G_i| + |G_j'|}  \in[0,1],\quad i=1,\ldots,K, \quad j=1, \ldots, K'. 
$$
with $|B|$ the number of elements in set $B$.\\
The similarity index is such that $\textrm{Sim}(G,G') \in [0,1]$ but it is not symmetric, therefore in order to have a symmetric measure we consider
$$
\overline{\textrm{Sim}}(G,G') = \frac12 \left(\textrm{Sim}(G,G') +\textrm{Sim}(G',G') 
\right).
$$
Figure~\ref{fig:allDays} is a heatmap representation of the matrix of similarity index among all MFAs for the period 1 February 2020 - 29 June 2020. Darkest-bluish zones are most different, whereas lighter-reddish are the most similar. It is interesting to observe the difference between before and after the lockdown (14 March 2020), then a slow recovery to normality. It is also worth noticing that holidays and weekends have clearly different mobility patterns than weekdays and that these MFAs are different from the administrative borders (Austrian districts), especially during weekdays. 
To put in evidence the intra weekly patterns in Figure~\ref{fig:similarity} we calculate the similarity  of the MFAs for the same day of the week. We start from the first Monday of the data and we calculate the similarity of it with all the subsequent Mondays. Similarly for the Tuesdays, etc.
We also calculate similarity of the daily MFAs with the administrative borders. What clearly emerges from Figure~\ref{fig:similarity} is that each day of the week has an almost stable pattern before and after the lockdown and in this sense the MFAs are similar among them for the same day fo the week. Further, we can notice that the similarity of the MFAs with the administrative borders is quite low, meaning that indeed MFAs do not coincide with the latter. The effect of lockdown is also seen in this curve.

\section{Detecting the persistent MFAs}\label{sec:persistent}
\begin{figure}[!htb] 
    \centering
        \includegraphics[width=0.45\textwidth]{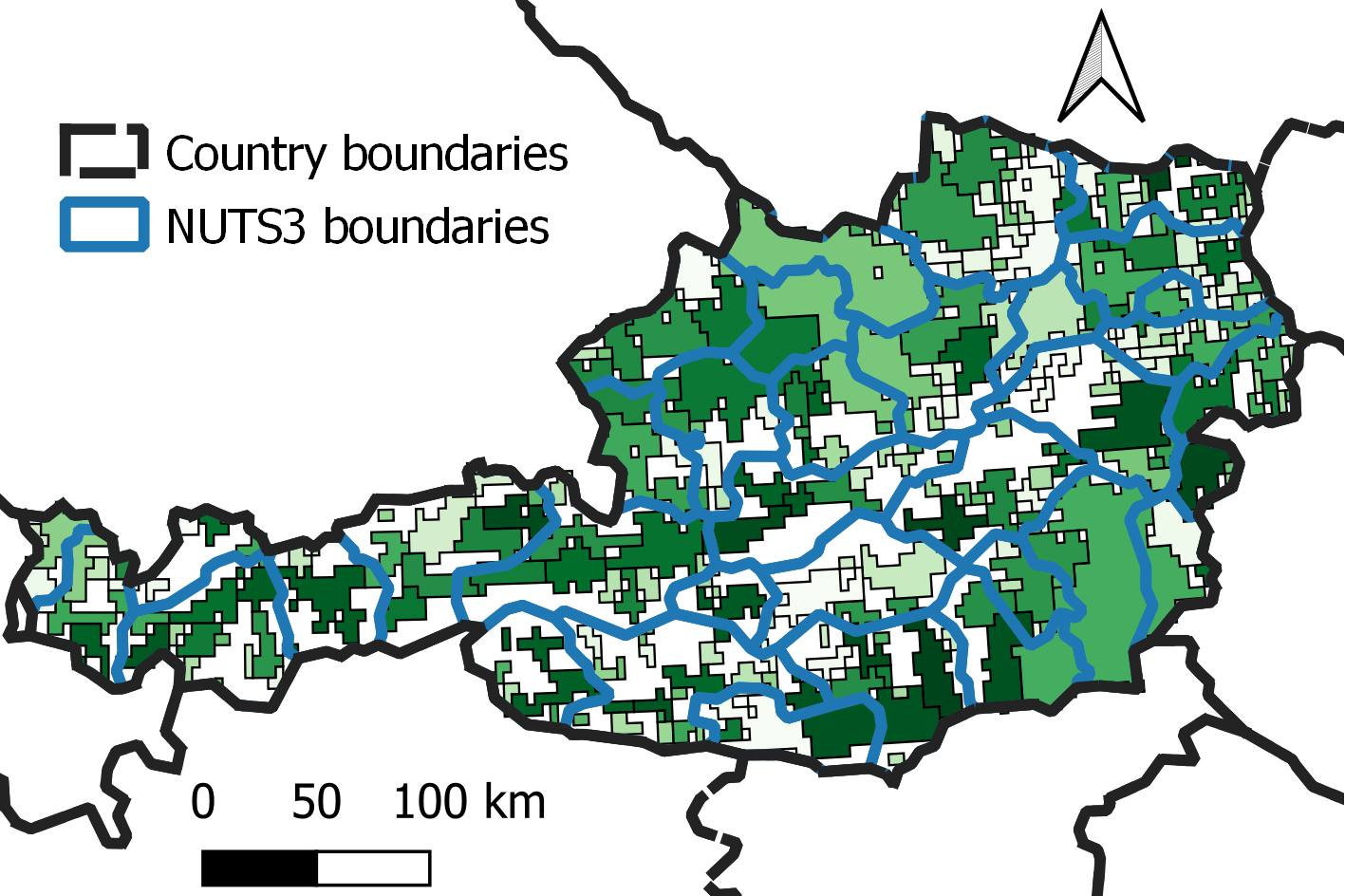}
        \includegraphics[width=0.45\textwidth]{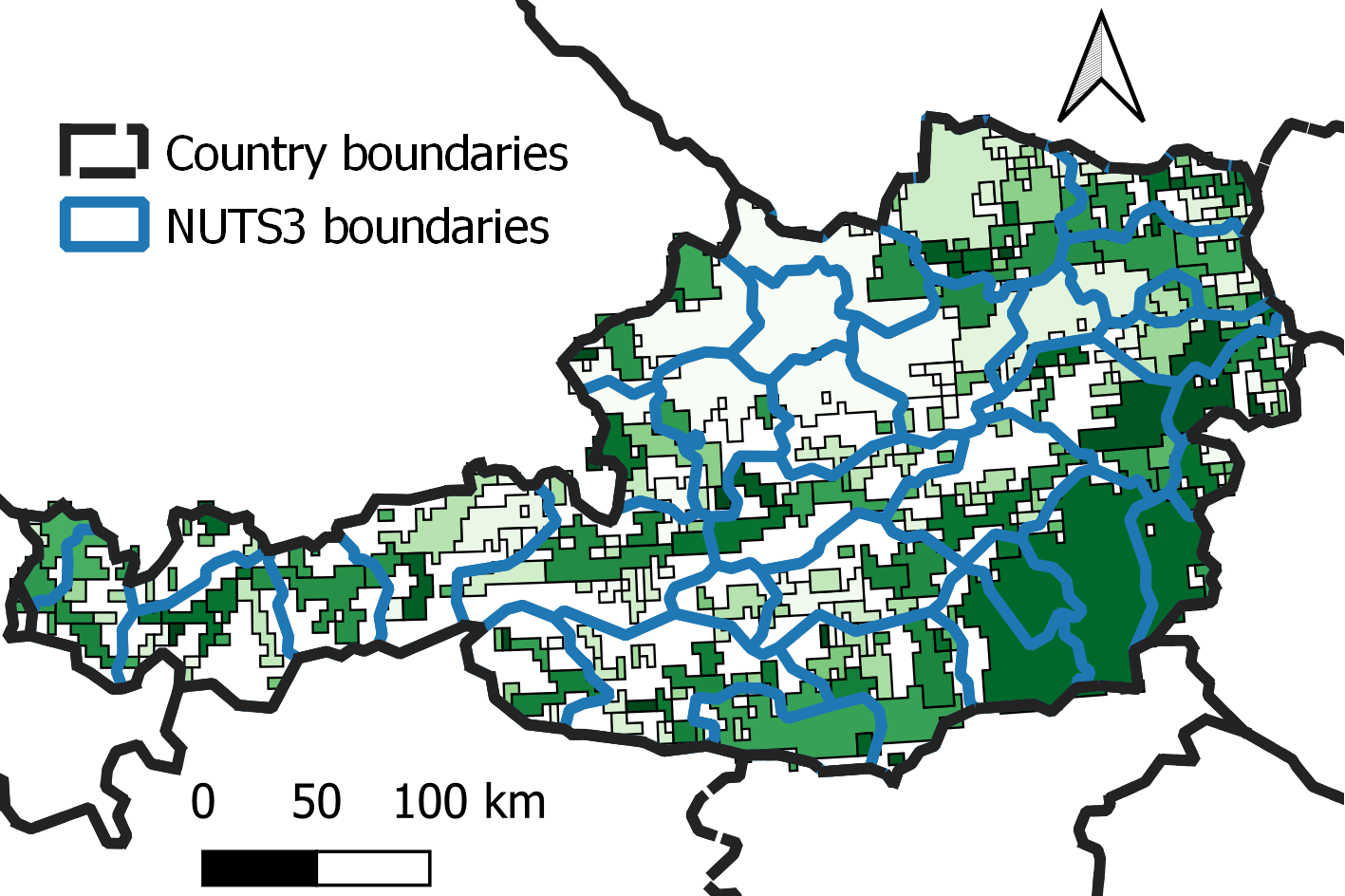}
\caption{Persistent and lockdown MFAs for Austria compared to NUTS3 districts (blue lines).   Whereas before the lockdown the MFAs extend across provinces, after the lockdown their area generally reduces and  they mostly lay within districts' borders (with some exceptions) or they completely disappear as all mobility remains within individual areas. Areas where connectivity is below 15\% (no apparent stable direction is observed in the mobility flows) are transparent. The color scale is only meant to distinguish the MFAs. Colors of the left and right maps are different as the number of MFAs is different.}
    \label{fig:persistent}
\end{figure}

As seen in the previous section, the MFAs have daily patterns, they change between before and after the lockdown is in force and tend to go back to their original shapes after the ease of containment measures. Moreover, MFAs shows time-variability also for the same weekday; thus, in order to fully exploit their potential, a stable version of the MFAs needs to be identified. Since the number of MFAs changes day by day and the same
 area may move from an MFA to another (changing the MFA label associated to it), we apply a $CO$-association method.
The $CO$-association method ($CO$) avoids the label correspondence
problem. It does so by mapping the ensemble members onto a new representation where the similarity matrix is calculated between a pair of objects in terms of how many times a particular pair is clustered together in all ensemble members \citep{FredJain2005}. In other words, $CO$ calculates the percentage of agreement between ensemble members in which a given pair of objects is placed in the same MFA.\\
As the first lockdown in Austria was enforced on 16 March 2020, we focus on the weekdays from 1 February 2020 to 15 March 2020.\\
Let $d$ be the data of one of these $D$ weekdays and $\textrm{MFA}_d$ the set of mobility functional areas obtained on day $d$. We then evaluate the \textit{co-association matrix} 
$$
CO(x_i,x_j) = \frac{1}{D}\sum_{d=1}^D \delta\left(MFA_d(x_i), MFA_d(x_j)\right)
$$
where $X_i$ and $x_j$ are the  areas  and $MFA_d$ is the set of MFAs for day $d=1, \ldots, D$ and $\delta(\cdot,\cdot)$ is defined as follows:
$$
\delta(u,v)=\begin{cases}
1,& \textrm{if } u \textrm{ and } v \textrm{ belong to the same MFA}, \\
0,& \textrm{otherwise}.
\end{cases}
$$
Then, as our scope is to obtain a \textit{persistent} version of the MFAs, we further threshold the $CO$ matrix so that all entries below 50\% are set to 0 and those higher or equal 50\% are set to 1 (it means that  only cells falling in the same MFA at least 50\% of the times are associated with that MFA) leading to a new matrix $\overline{CO}$.\\
Then again a directed graph is built with this matrix using the entries of the $\overline{CO}$ matrix to weight the edges and applying the walktrap algorithm to obtain the final persistent MFA.
The same procedure is replicated for the lockdown dates, ending up with a different set of  stable  MFAs that we denote by  lockdown-MFA.
With these two sets of MFAs at hand, we further test if they are meaningful to the analysis. It turns out that these stable MFA are in fact reasonably well defined.\\
We then apply the symmetric similarity index $\overline{\textrm{Sim}}(\cdot,\cdot)$ for all the daily MFAs against the persistent MFAs, the lockdown MFA and the districts (NUTS3). Figure~\ref{fig:persistent}.

\begin{figure}[!htb] 
    \centering
    \includegraphics[width=\textwidth]{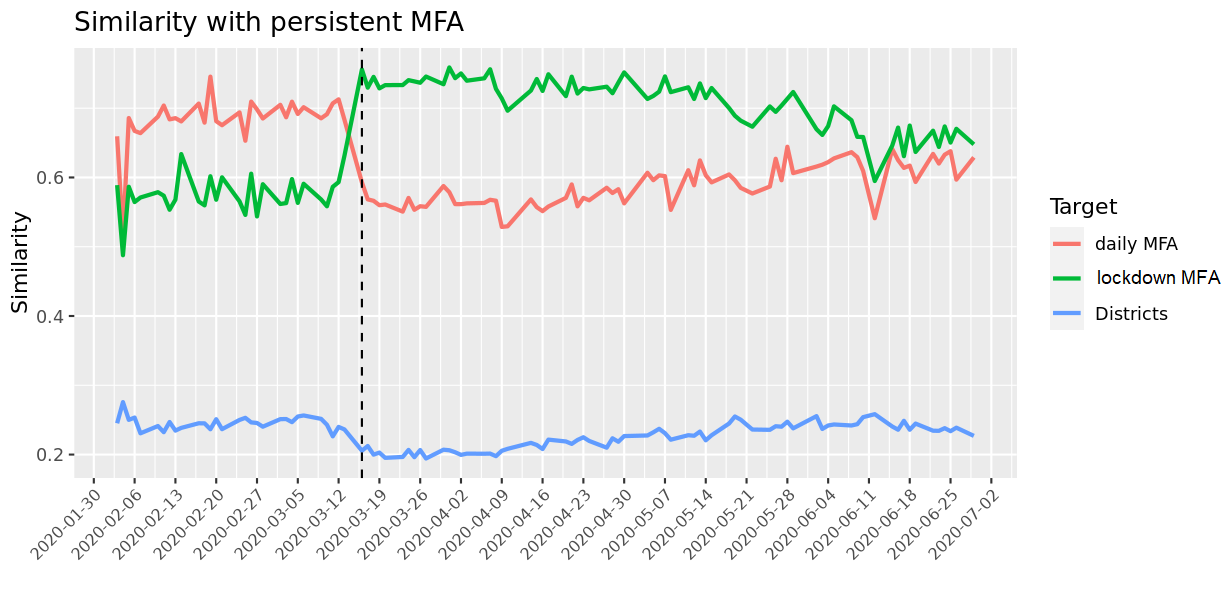}
    \caption{Similarity of daily MFAs (orange line), lockdown MFAs (green line) and Districts (blue line) with respect to persistent MFAs. Daily MFAs are very similar to the persistent MFAs before lockdown which was enforced on 16 March 2020, whereas they are more similar to lockdown MFAs after. A return to normality is slowly appearing. Once again, the MFAs are quite different from administrative borders.}
    \label{fig:similarityMFA}
\end{figure}

In Figure~\ref{fig:similarityMFA} we calculate the similarity index between the daily MFAs against the lockdown MFAs and, against the districts. It quite evident the abrupt change in the shape of the MFAs on March 16h which corresponds to the stay-at-home nation wide order imposed by the Austrian authorities.

\section{Potential applications of Mobility Functional Areas}\label{sec:Vienna}
The MFAs, originally developed in the framework of the non-medical interventions to the fight the spread of COVID-19 in Europe, have many other fields of application. In this section, we briefly take an overview of some of these fields.

\subsection{Transport system}
\begin{figure}[!htb] 
    \centering
    \includegraphics[height=0.3\textheight]{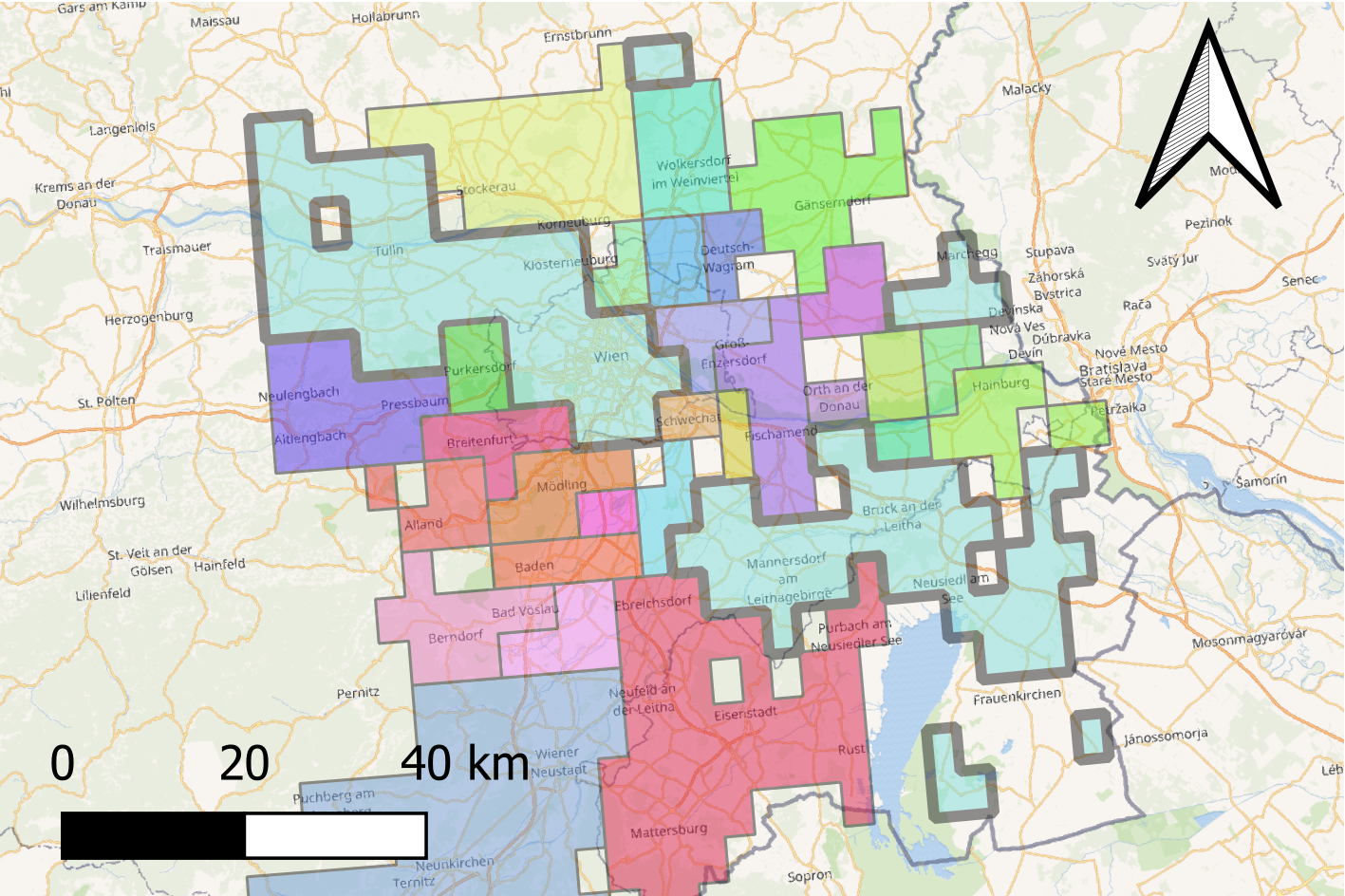}
    \includegraphics[height=0.3\textheight]{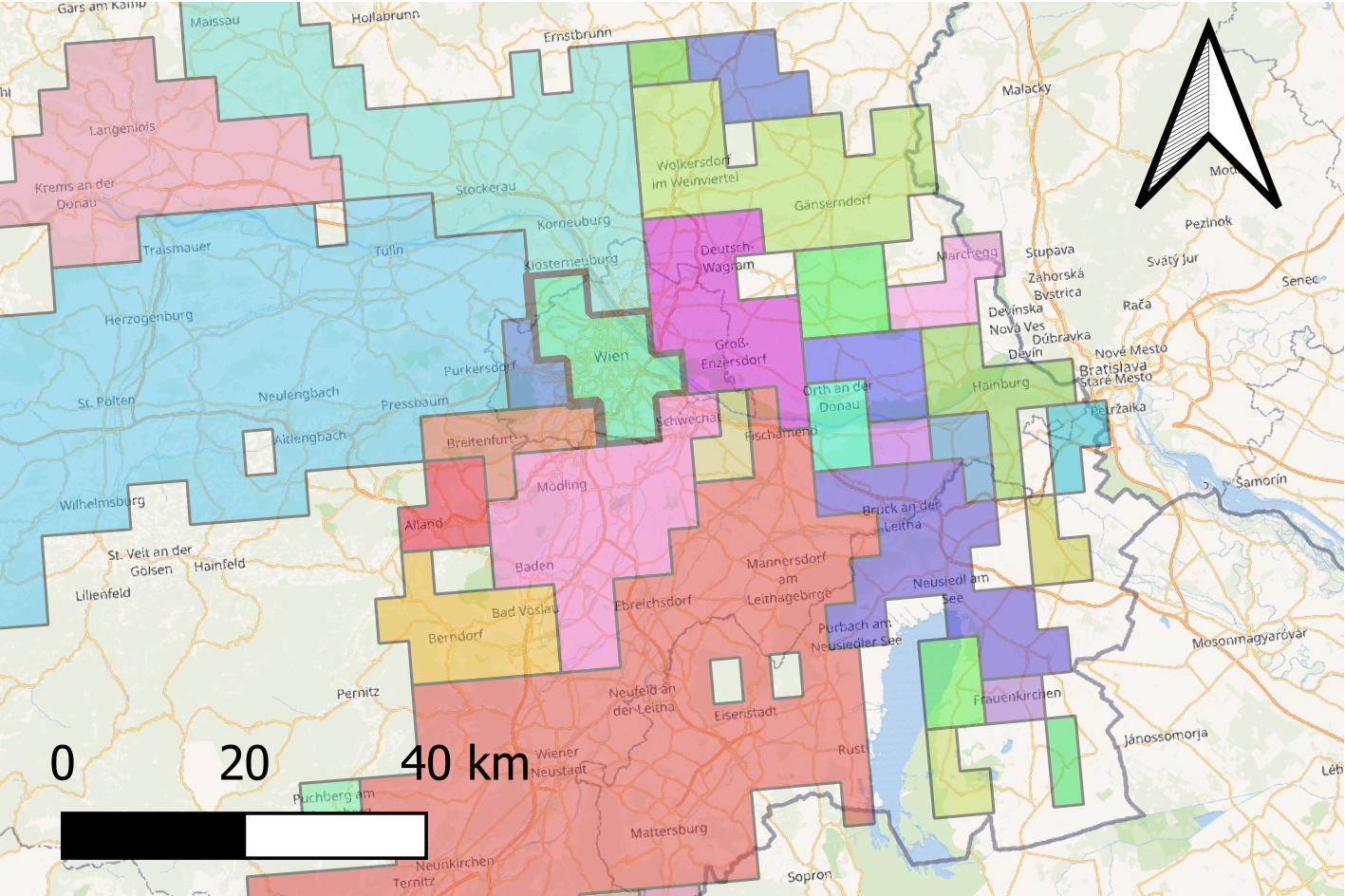}
    \caption{The persistent MFAs around Vienna before (top) and during the lockdown (bottom). Colors are randomly assigned to the MFAs, without any relation between top and bottom maps. The largest persistent MFA around Vienna (top map), marked with a bold gray border, is made of non contiguous zones and follows the main highways passing through the capital. This MFA is highly shrunk during lockdown (bottom map). On the other hand, some of the MFAs merge during lockdown and there are no MFAs made by non contiguous zones. This seems indicating that movements are likely to be more local during the lockdown.  Background map Open Street Map: Road network in orange; Country boundaries in grey; water surfaces in light blue}.
    \label{fig:Wien}
\end{figure}

As said, the MFAs are data-driven products generated by natural human mobility, which is in fact strictly related to available local transport system; this is indeed a bi-directional relationship. At the same time, the MFAs can inform about possible new directions in which the transport system can be enhanced. Another interesting aspect of the MFAs is  that, especially around big cities, they are not necessarily made of contiguous zones; although this is an artifact, or better a feature of the MFAs due to how they have been designed, starting form Origin-Destination Matrices (ODMs). This does not allow showing that when going from $\bf a$ to $\bf c$ one can stop by a bar to take a coffee in $\bf b$. For example, in Figure~\ref{fig:Wien} we can see some typical stylized facts of MFAs:
\begin{itemize}
    \item MFAs in some cases follow already existing infrastructures, like highways;
    \item MFAs are not contiguous;
    \item MFAs are not all of the same size and this is clearly related to the relative importance of the geographical zones involved.
\end{itemize}
The largest MFA in the top map of Figure~\ref{fig:Wien} is clearly made of contiguous and not-contiguous areas and its development follows the the most important highway crossing the capital. These zones, which are highly interconnected by definition of MFA, represent important highlights to eventually improve the transport system accordingly.
The bottom map in Figure~\ref{fig:Wien} shows how the MFAs have shrunk during the lockdown.\\

Following the same approach used for the Vienna example, but outside the pandemic context, MFAs calculated over two different time periods and compared with each other can also help monitoring the actual usage of transport infrastructures in order to support their management.
Moreover, by analysing daily MFAs, local administrations can monitor how mobility changes in occasion of major mass-gathering events, learning how to plan local mobility restrictions and alternative paths.

\subsection{Socio-economic impact of the pandemic}
By comparing the two maps in Figure~\ref{fig:Wien}, it is evident how MFAs can show major changes in human mobility patterns. Eventually, with MFAs built starting from higher spatial-resolution data, it would have been also possible to asses the socio-economic impact of the lockdown measures at urban level. As an example, one would have been be able to identify neighborhoods where the residential population kept moving for working-related reasons (because they could not telework or they were essential workers), despite the mobility restrictions and the risk of being infected. Another example is the ability to monitor whether or not key commercial areas (such as malls, shopping hubs, etc.) had remained connected in terms of mobility with the high-population-density neighborhoods.

\subsection{Public health}
The MFAs were originally designed to support the fight to the spread of COVID-19 in Europe. Their role in the public health sector, specifically during a pandemic, is therefore evident and is further developed in Section~\ref{sec:austria}. Beside their employment in an emergency context, MFAs could be also exploited to plan and monitor the accessibility of local health units, main hospitals and any other type of health facility.

\subsection{Urban planning}
The persistent MFAs can help authorities and spatial planning agencies to take more evidence-based and informed decisions. Sustainable spatial planning and spatial management requires the identification of the optimal location for both private and public facilities and services that minimize transportation needs while maximizing service availability. For example, MFAs can assist spatial planners and developers on identifying the optimal location of new housing developments, schools and sports facilities.

\subsection{Environmental risk and pollution}
Since the MFAs show human mobility, when it comes to environmental monitoring and risk management, they can be profitably used to design the displacement of monitoring stations, in order to measure those pollutant associated to road-mobility. Analogously, MFAs could support the planning of new waste-to-energy plants, wastewater treatment plants biomass plants, etc, outside the main human-mobility areas.  

\subsection{Demography and Migration}
MFAs can be used to identify disadvantaged/isolated areas which are not well connected to economically developed MFAs. These areas are more likely to experience an aging population as well as a depopulation. Besides, without any administrative intervention, these areas are likely to be less attractive to international migrants \cite{demographicjrc2021}.

\section{MFAs and COVID-19 cases during the first wave}\label{sec:austria}
\begin{figure}[!htb] 
    \centering
    \includegraphics[width=0.9\textwidth]{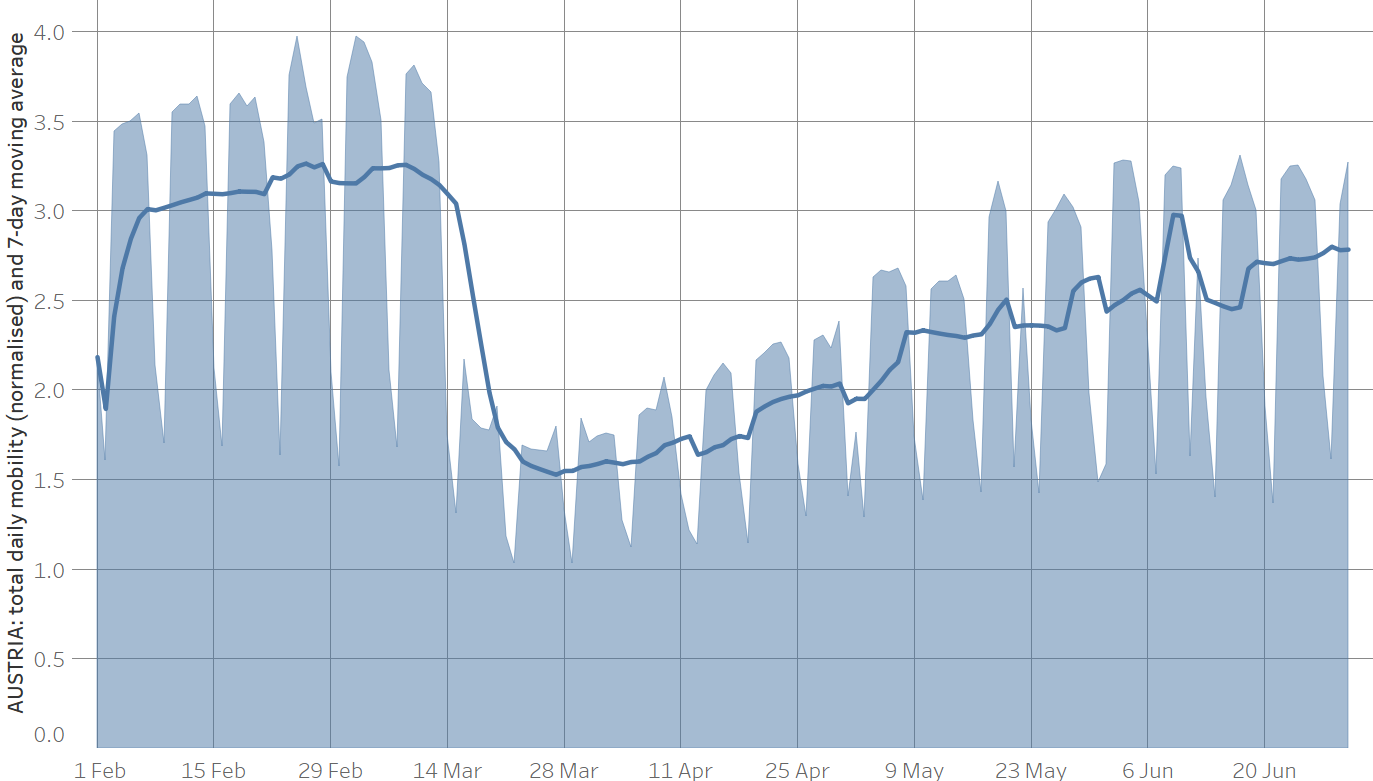}
\caption{Relative mobility change in Austria during the period 1 February 2020 - 30 June 2020.}
    \label{fig:AustriaMNO}
\end{figure}

In this section, we analyse the role of MFAs in the contest of the spreading of COVID-19.
The objective is testing the hypothesis that MFAs are correlated to the evolution of the number of cases until strict measures to limit mobility are enforced. For this reason, we focus on the period between 15 March 2020 and 30 April 2020, monitoring the number of cases in the areas laying within or outside the persistent MFAs. 
It is worth noticing that the actual timeline of the measures taken is the following\footnote{Source Wikipedia: \url{https://en.wikipedia.org/wiki/COVID-19_pandemic_in_Austria}}: a) On 13 March 2020 the entire valley surrounding Ischgl was in lockdown; b) On 15 March the state of Tyrol was in lockdown; c) From 16 March a nationwide lockdown was implemented in Austria. Thus given the limited period of time that local measures were in place we do not expect any endogeneity issues in the statistical analysis that follows. Figure~\ref{fig:AustriaMNO} shows the relative change in mobility for Austria during the period 1 February 2020 - 30 June 2020.

\begin{figure}[!htb] 
    \centering
    \includegraphics[width=0.3\textwidth]{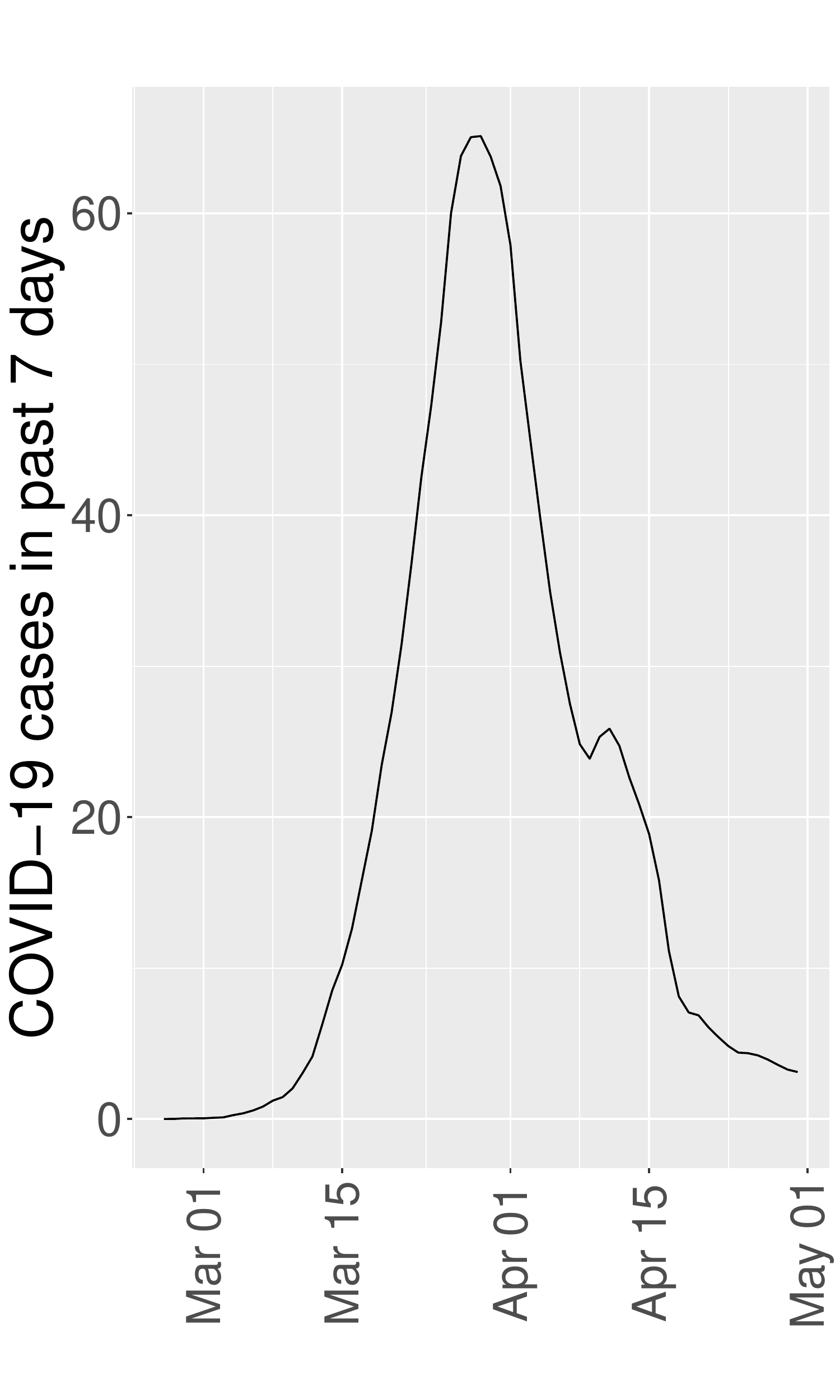}
        \includegraphics[width=0.3\textwidth]{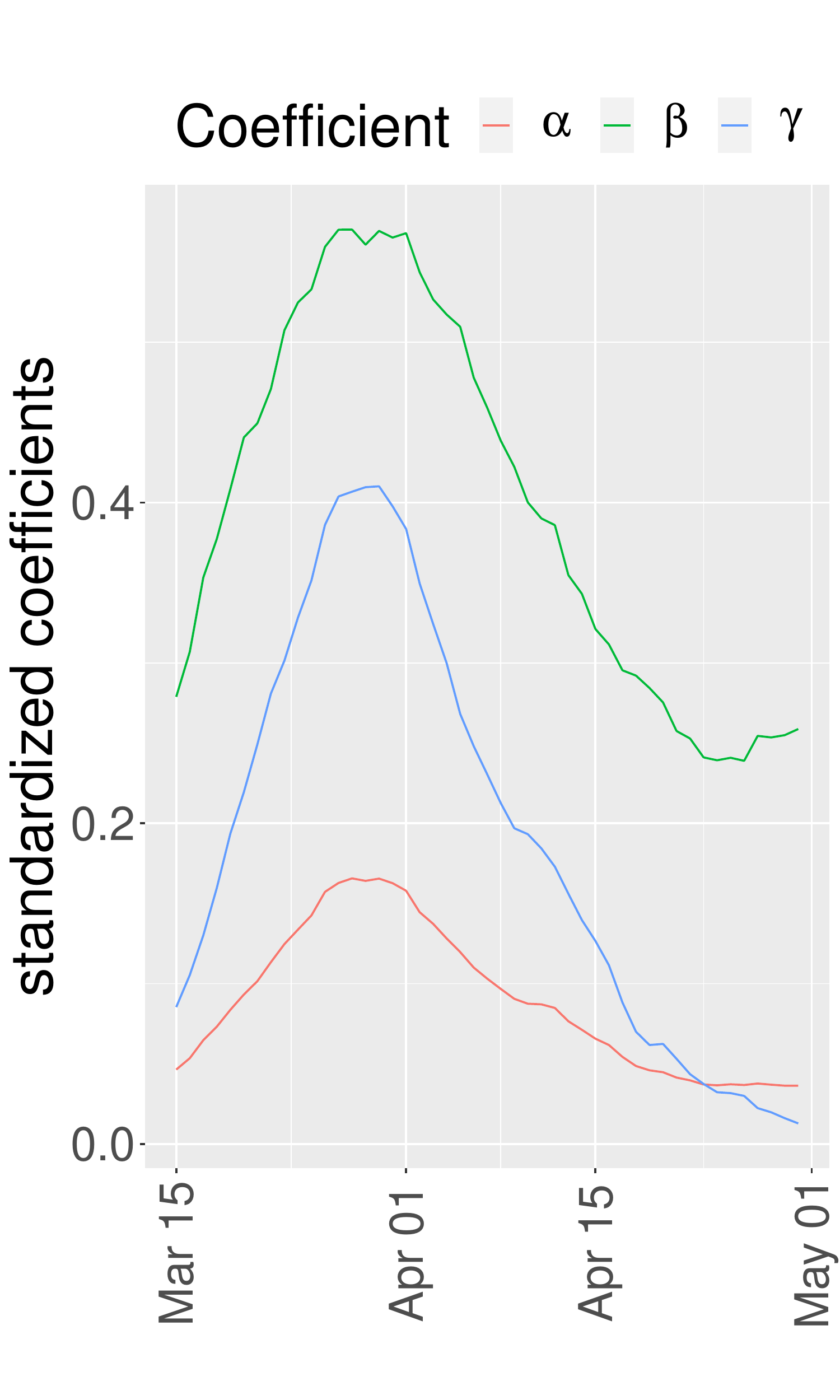}
        \includegraphics[width=0.3\textwidth]{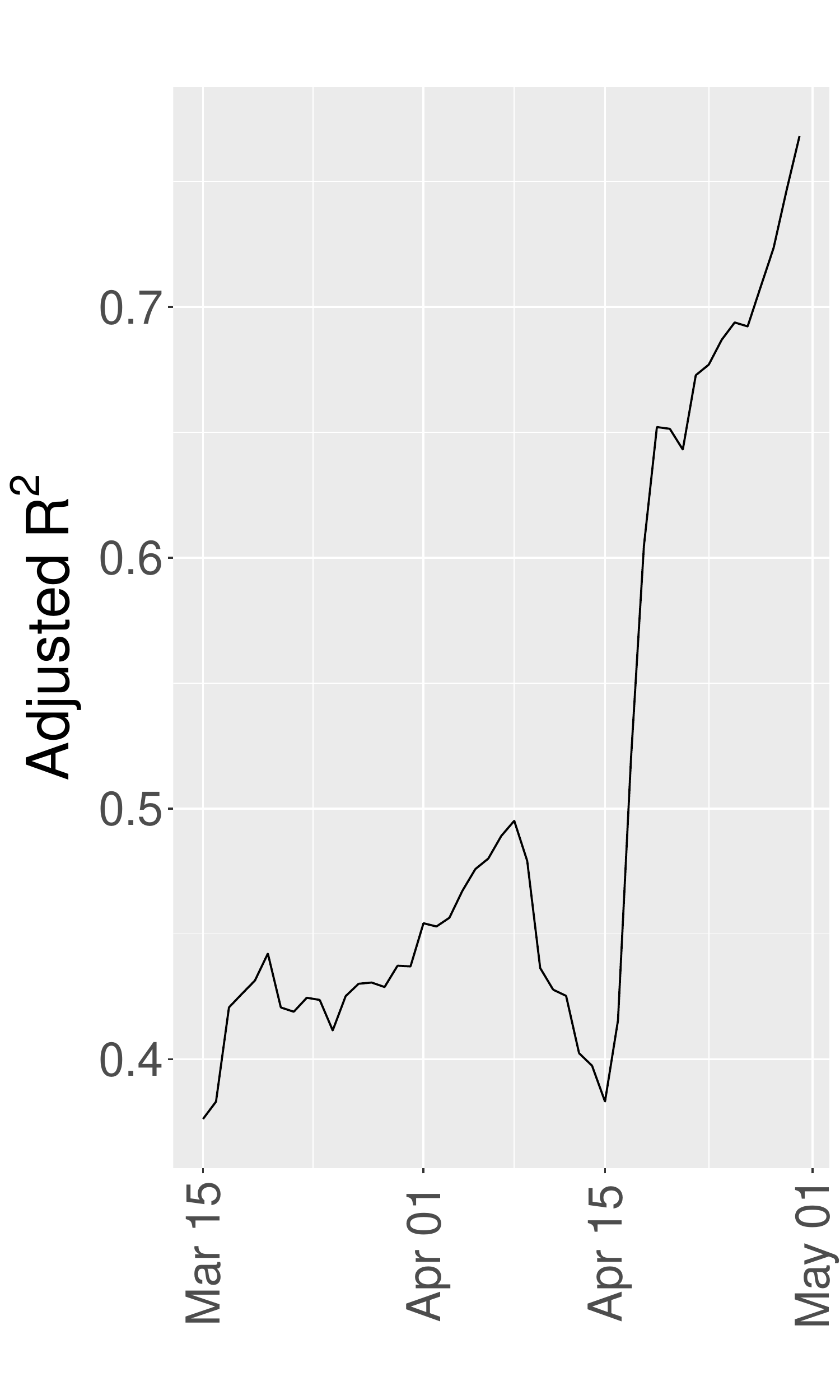}
\caption{Left to right: the number of cases in past 7 days, the value of the coefficients $\alpha$, $\beta$ and $\gamma$ in model \eqref{eq:mod1} and the adjusted $R^2$ of the estimated model \eqref{eq:mod1} for each date. Coefficients are always significant at level 0.01.}
    \label{fig:Coeff}
\end{figure}

It is known in the literature that it takes about 14 days \citep{TR-Connectivity} to see the decrease in the number of cases due to a lockdown, so we should expect a peak around the 30th of March and then a decrease of cases as indeed Figure~\ref{fig:Coeff} shows.

\begin{figure}[!htb] 
    \centering
    \includegraphics[width=0.9\textwidth]{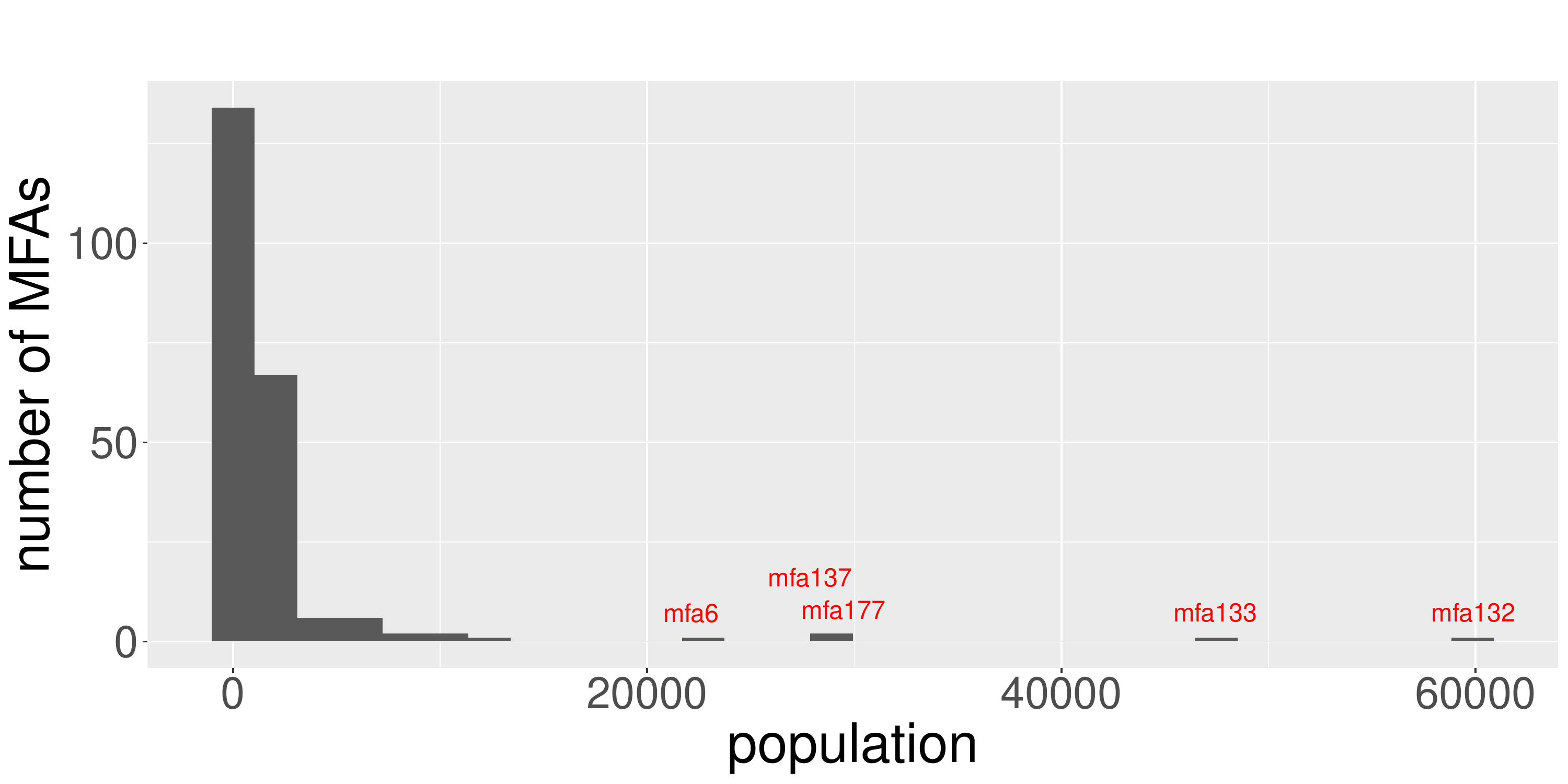}
\caption{Distribution of population by MFA. Only 5 MFAs are large, i.e. more than 20000 inhabitants, in population size and all but one overlaps with Wien district (see also Table~\ref{tab:MFApopDist}).}
    \label{fig:mfaPopDist}
\end{figure}

\begin{figure}[!htb] 
    \centering
    \includegraphics[width=0.45\textwidth]{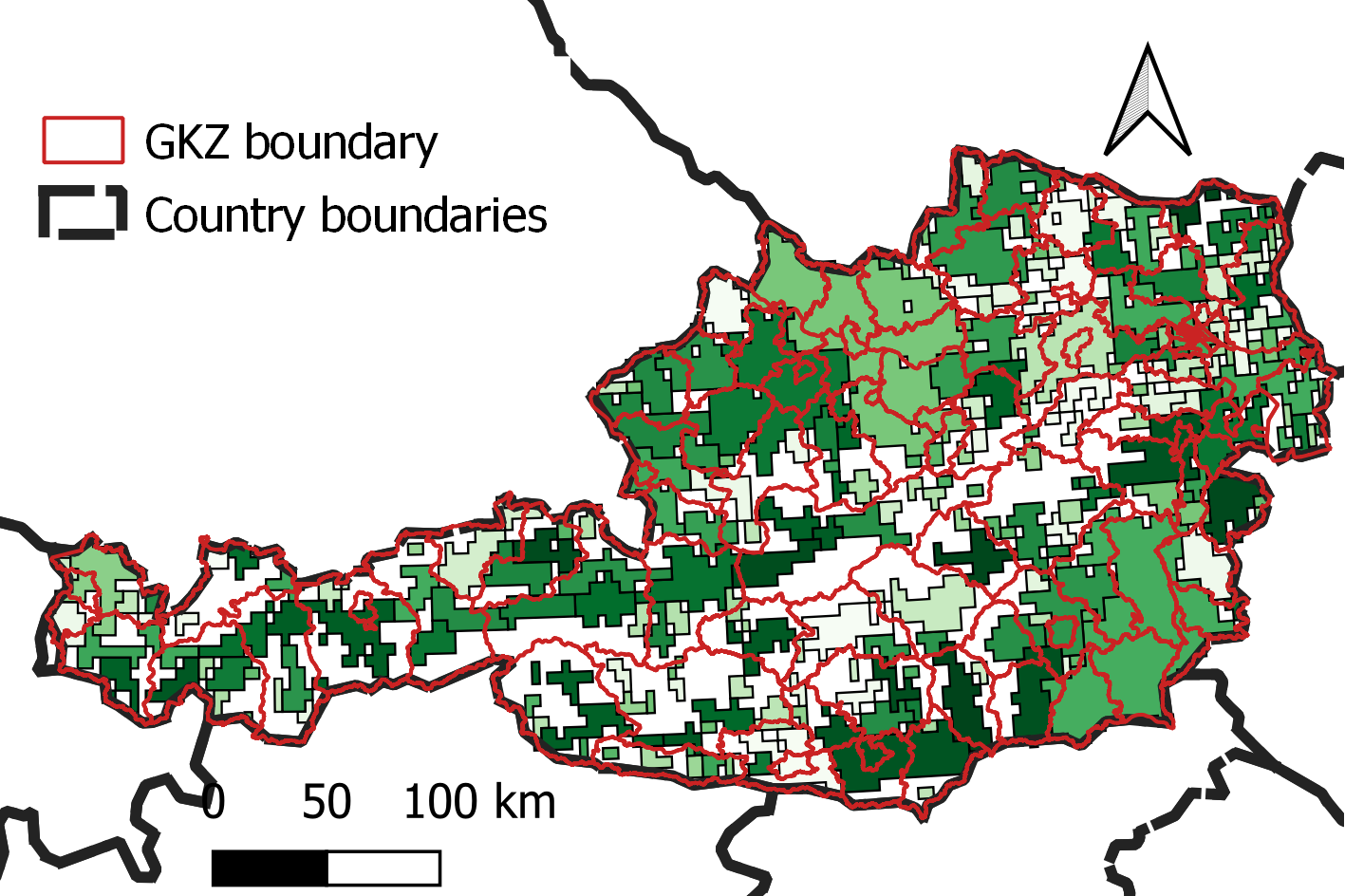}
    \includegraphics[width=0.45\textwidth]{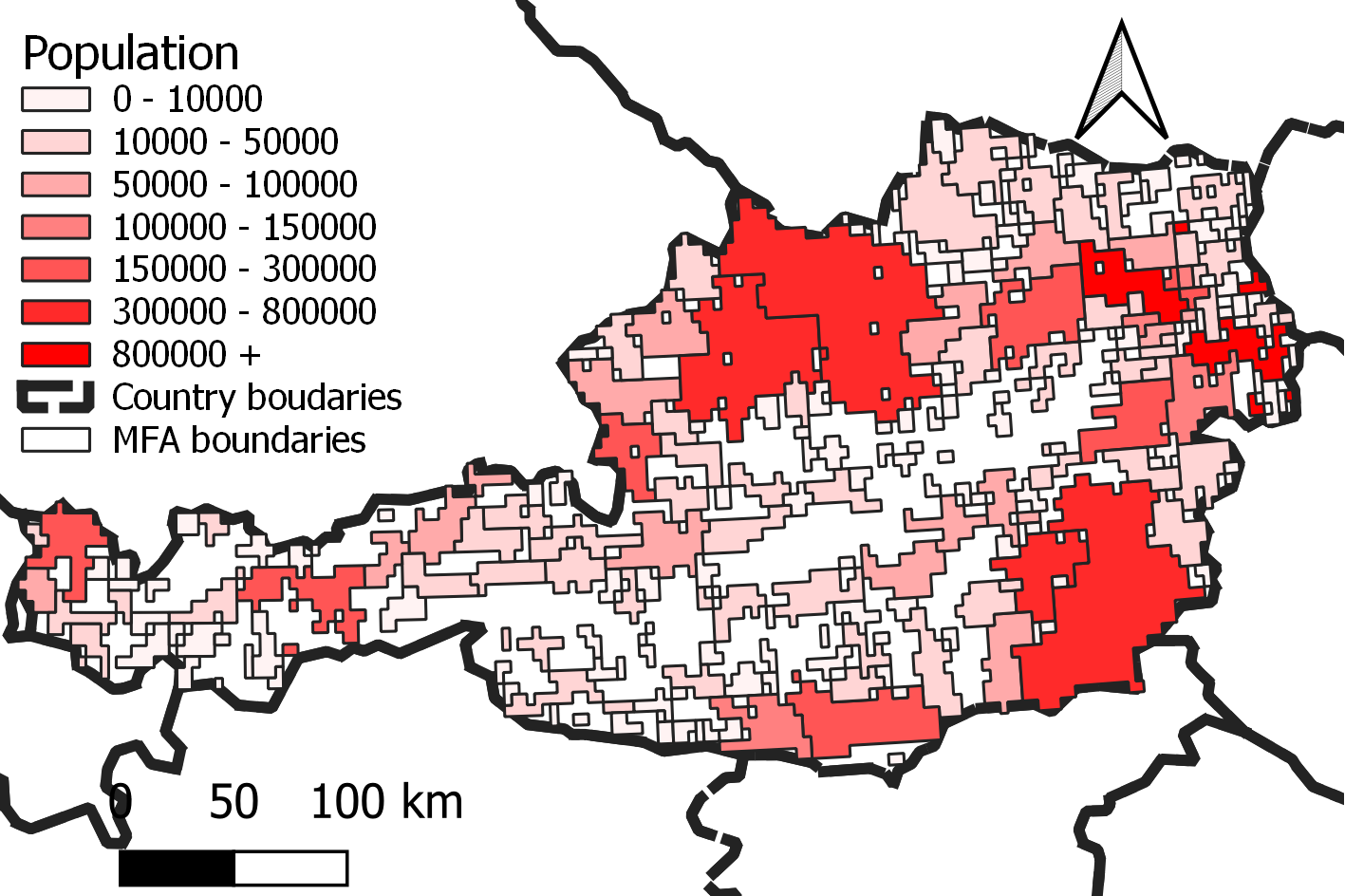}
\caption{MFAs compared to GKZs borders (left) and the estimated population on the MFAs (right). The MFAs also cross the administrative borders on which the health data are available.}
    \label{fig:mfaPop}
\end{figure}

In Austria we have 222 persistent MFAs (plus the residual territory) built on 4789 areas identified by the MNO, while the number of political districts (GKZ) is 94. The GKZ do not match with the NUTS3 districts used in Sections~\ref{sec:MFA} and \ref{sec:persistent} and these are the geographical units for which epidemiological data is available.
Of the 4789 zones, 3409 belong to a proper MFA and 1380 are in the residual territory of Austria.
Not all the MFAs are big enough to be relevant either because their population is small or because their territorial extent is very limited. Figure~\ref{fig:mfaPopDist} shows the distribution of MFAs by population size and Table~\ref{tab:MFApopDist} shows the corresponding intersecting districts. As it can be seen, in all but one case, these MFAs contain the territory of Wien. Table~\ref{tab:MFAsize} reports the extension of the MFAs in $km^2$.
Using high resolution Facebook population data\footnote{\url{https://dataforgood.fb.com/tools/population-density-maps/}}, we reconstruct the population of each of 4789 the zones. Figure~\ref{fig:mfaPop} shows the shapes of the MFAs compared to those of the political districts for which health data are available and also the estimated population for each zone of the MFAs.
\begin{table}[!htb]
    \centering
\begin{tabular}{c|r|r|l}
MFA&Population&$km^2$&Districts that intersect the MFA\\
\hline
6         & 22126 & 96&M\"{o}dling, Sankt Pölten(Land), Wien(Stadt)\\
137         & 27870& 64&Korneuburg,  Wien(Stadt)\\
177         & 29479& 32&Bruck an der Leitha\\
133         & 47442& 80 &G\"{a}nserndorf, Wien(Stadt)\\
132         & 59897& 48&Wien(Stadt)\\
    \end{tabular}
    \caption{The MFAs with more than 20000 inhabitants. See also Figure~\ref{fig:mfaPopDist}.}
    \label{tab:MFApopDist}
\end{table}
\begin{table}[!htb]
    \centering
\begin{tabular}{l|c|c|c|c|c|c}
MFA type &Min. &1st Qu. & Median  &  Mean& 3rd Qu.    &Max.\\ 
\hline
persistent    & 16     & 48    &  89 &    248  &   246 &   4765\\
lockdown    & 31&      33     & 65 &    159&     143   & 2302
    \end{tabular}
    \caption{Distribution of the areas of the MFAs in $km^2$ before (persistent) and during the lockdown.}
    \label{tab:MFAsize}
\end{table}

It is also worth to mention that the daily density of movements\footnote{By daily density of movements we mean the average number of movements per day, in the pre-lockdown period, per cell of the grid that defines each MFA.} in the \textit{residual} MFA is equal to 0.00012, while on average   the other MFAs have a density of 15.016 (Q1 = 4.065, median = 8.100, Q3 = 13.680, max  = 294.111), meaning that most of the mobility is observed in the identified MFAs. Notice further that the residual MFA is made of 1553 cells of the grid, while on average an MFA is made of 14 cells with a maximum of 279 (Q1=3, Q3=15).

\begin{figure}[!htb] 
    \centering
    \includegraphics[width=0.95\textwidth]{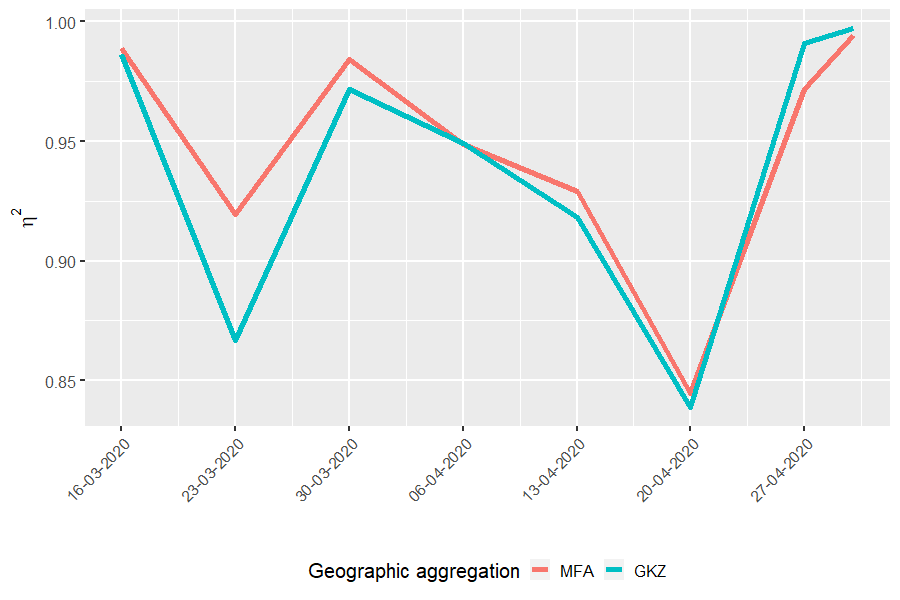}
\caption{The index  $\eta^2_{Y|G}$ by calendar week by group $G$  = MFA or $G$ = GKZ. The higher the value of $\eta^2_{Y|G}$, the more $Y$ is determined by $G$. Notice that $\eta^2_{Y|MFA}\geq \eta^2_{Y|GKZ}$ till 20 April 2020, then the reverse is true. This is expected because few weeks after the national lockdown, the impact of the MFA does no longer appear as the mobility has been reduced or stopped completely.}
    \label{fig:eta2}
\end{figure}

\begin{figure}[!htb] 
    \centering
    \includegraphics[width=0.45\textwidth]{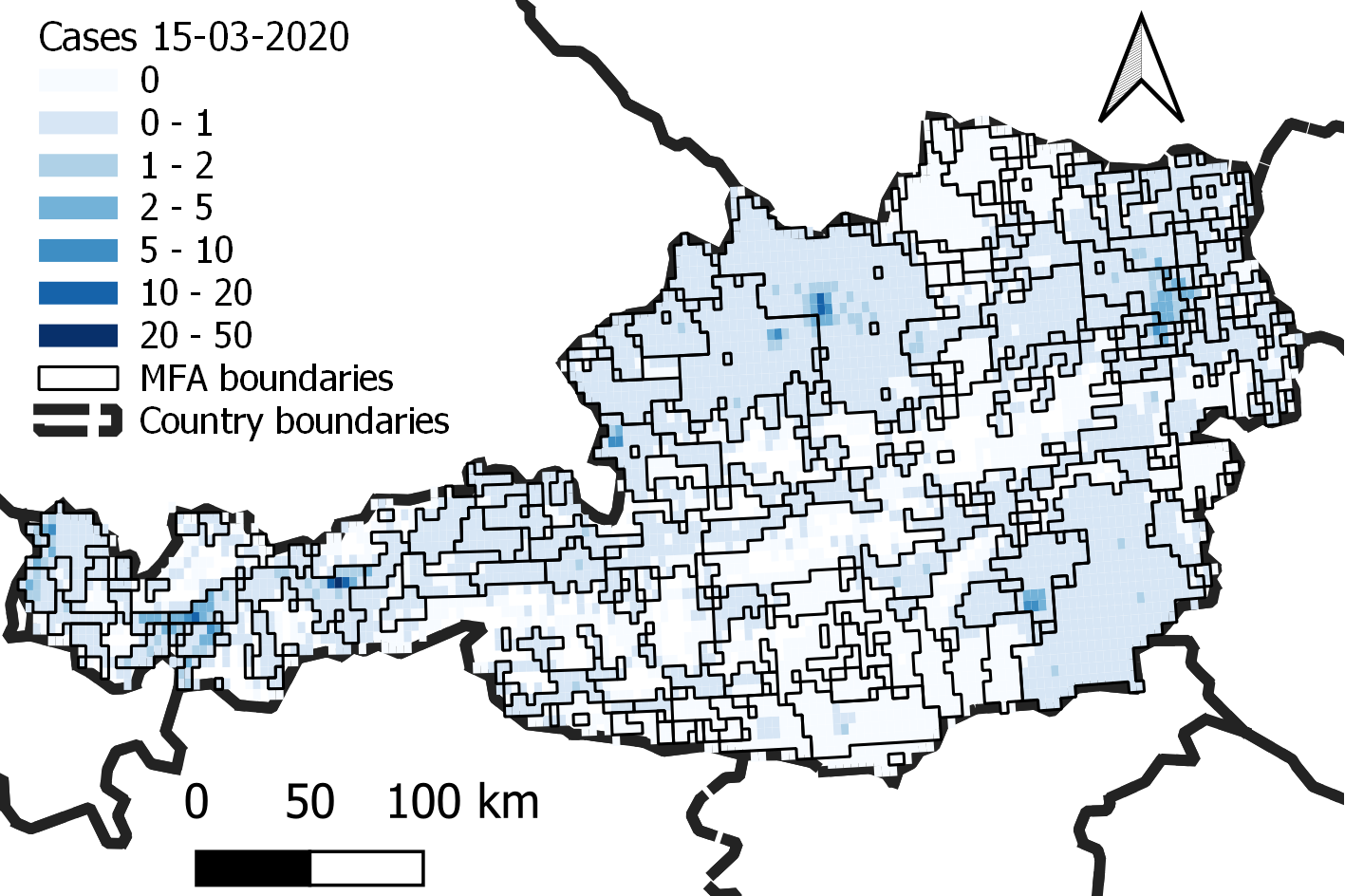}
    \includegraphics[width=0.45\textwidth]{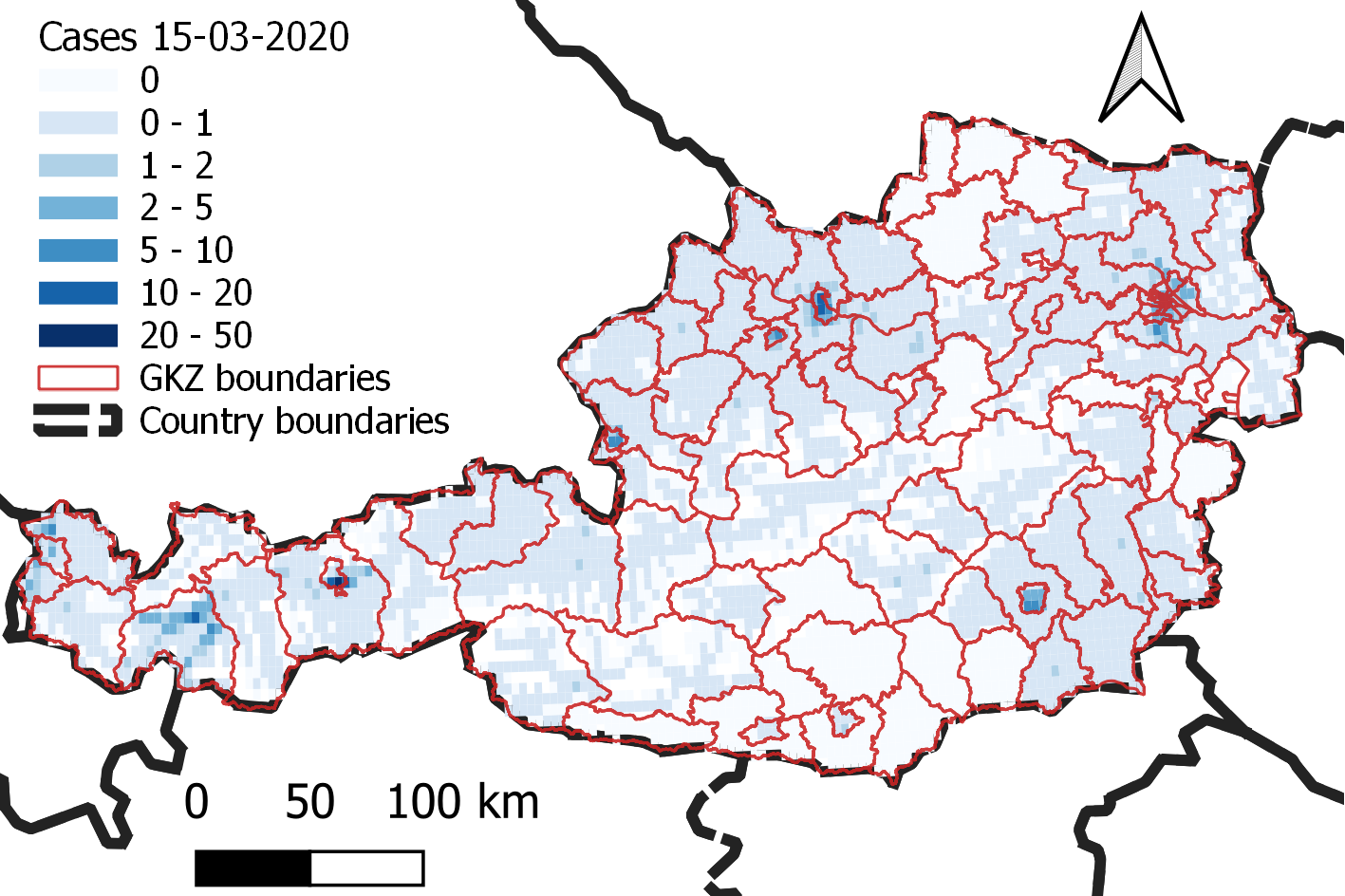}
        \includegraphics[width=0.45\textwidth]{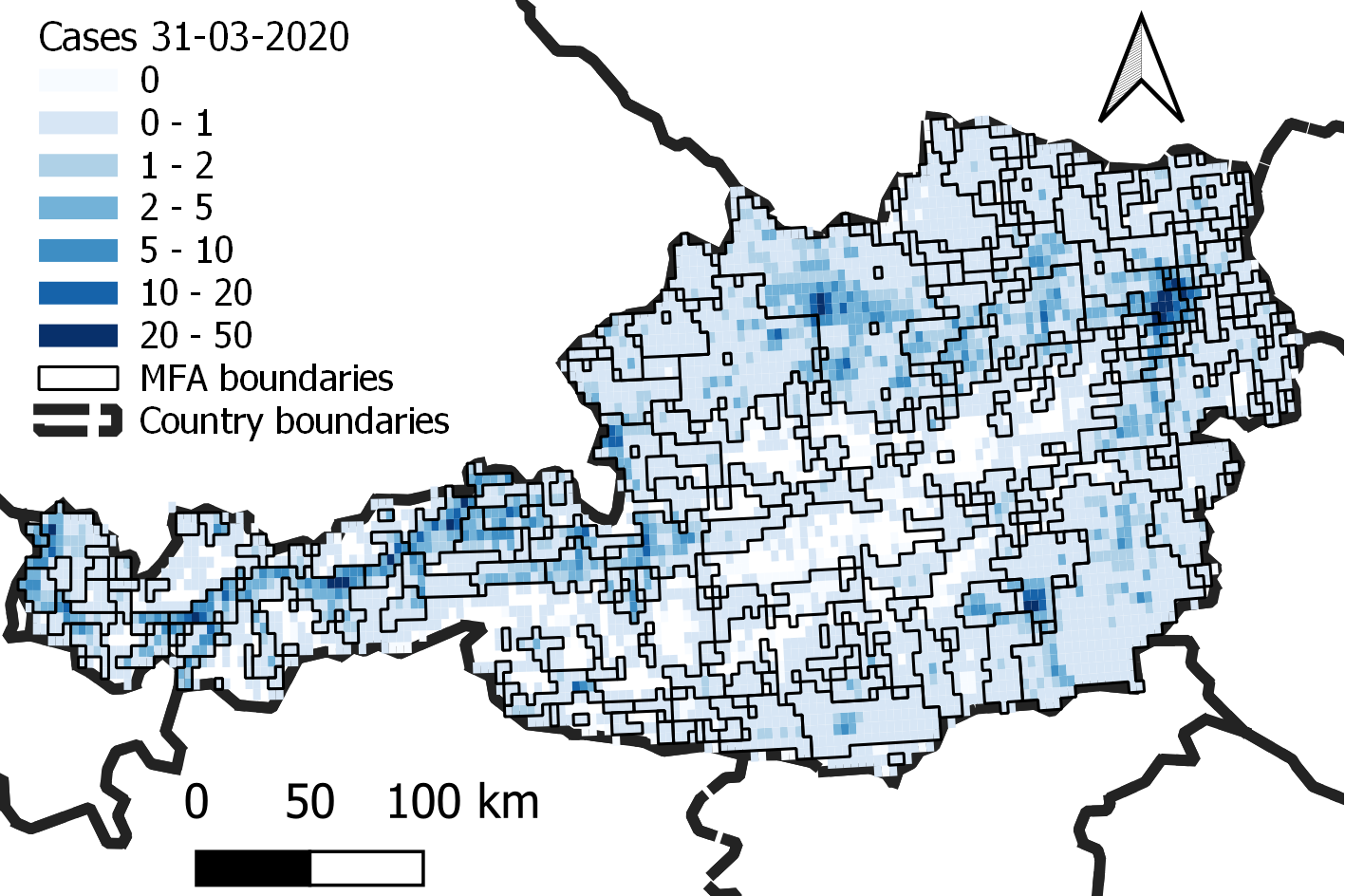}
        \includegraphics[width=0.45\textwidth]{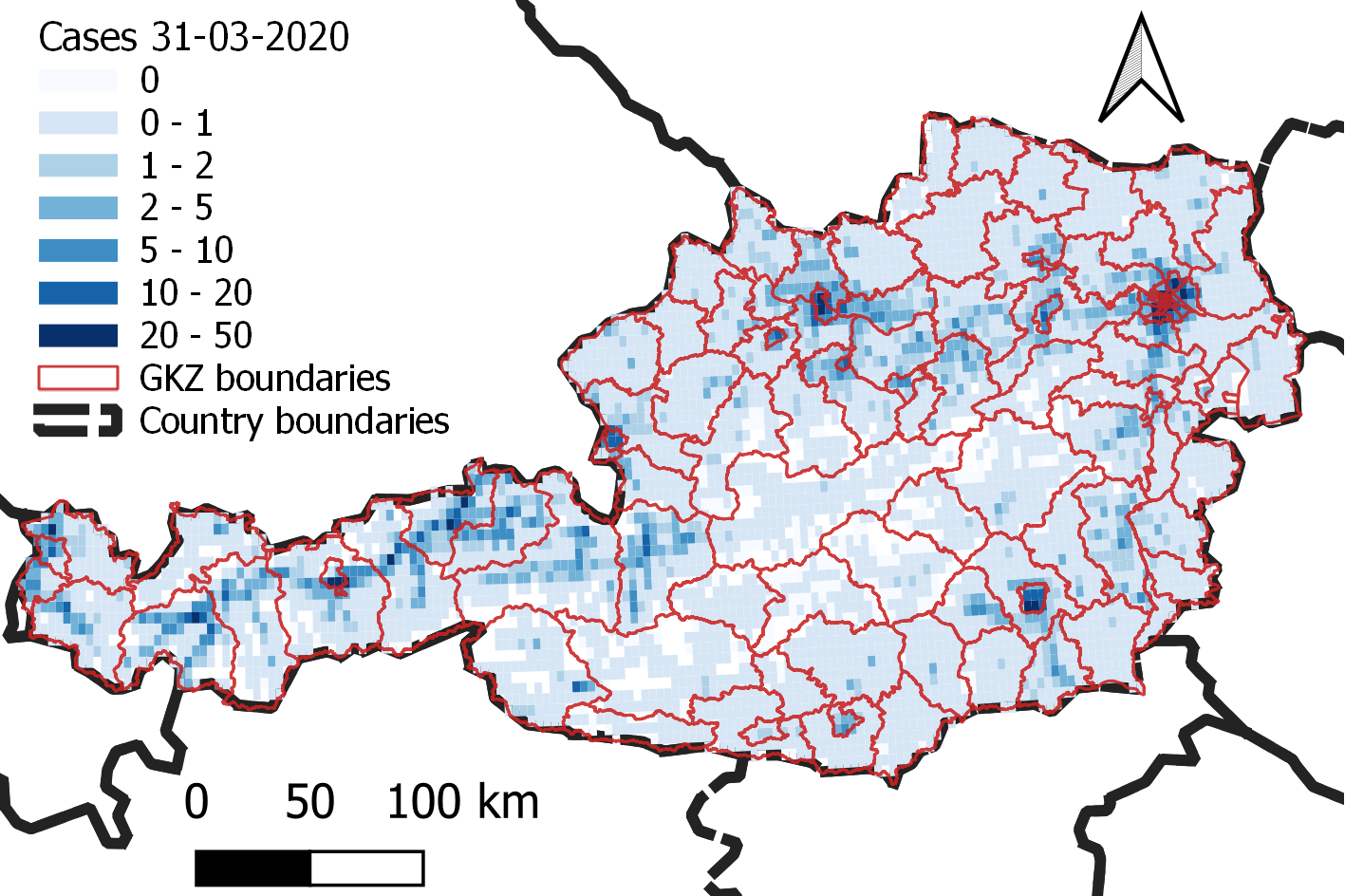}
        \includegraphics[width=0.45\textwidth]{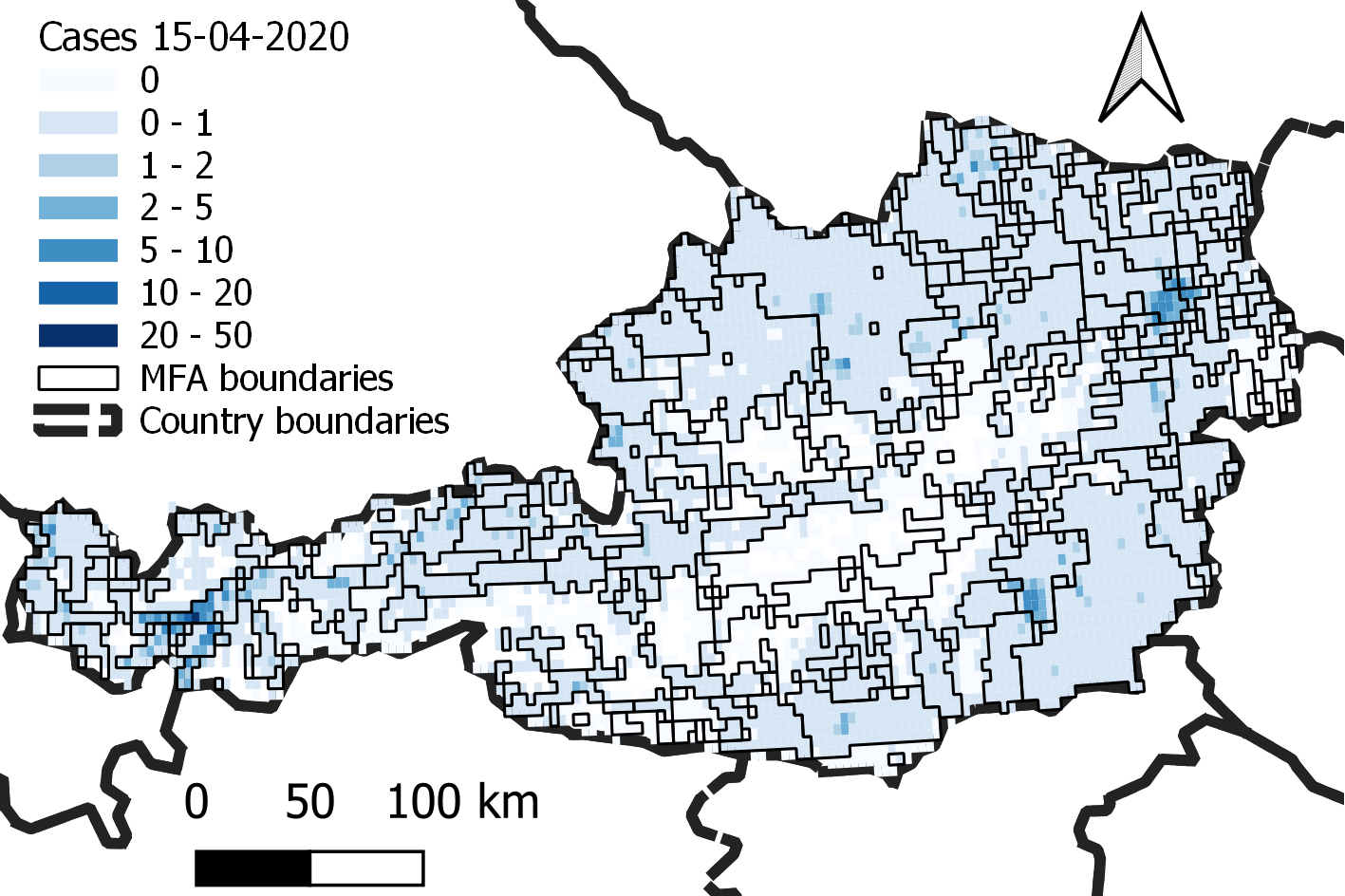}
        \includegraphics[width=0.45\textwidth]{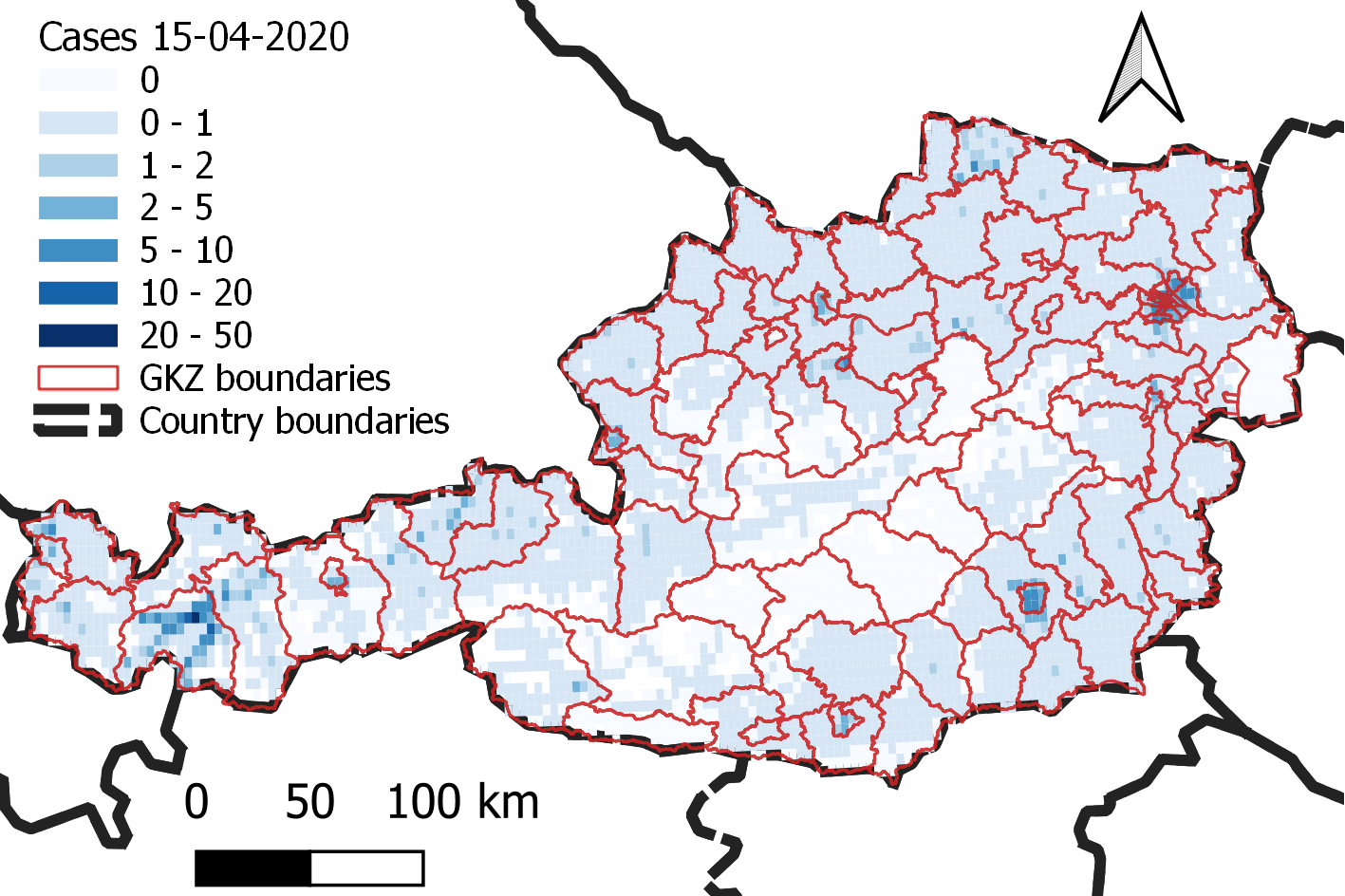}
        \includegraphics[width=0.45\textwidth]{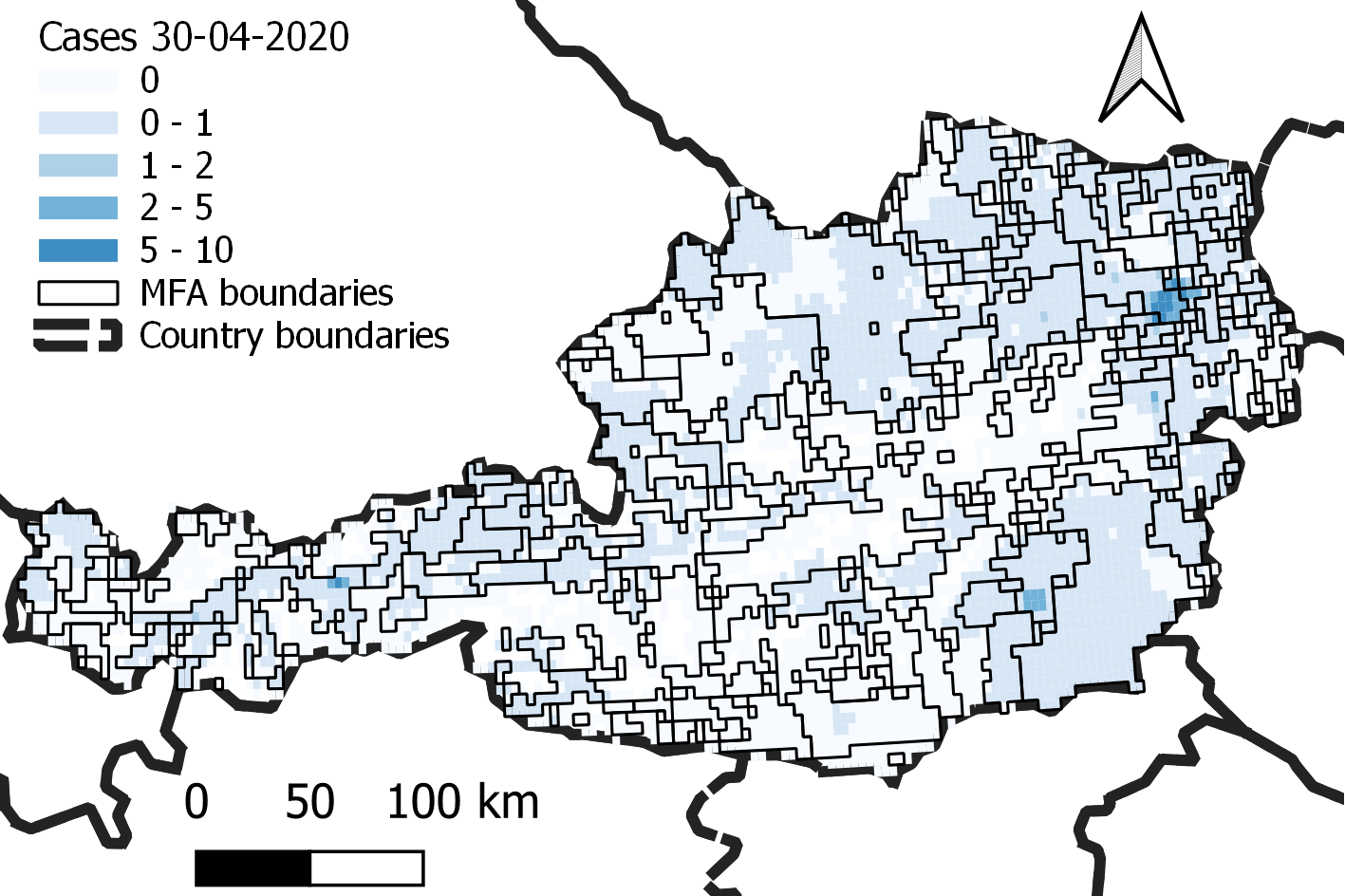}
        \includegraphics[width=0.45\textwidth]{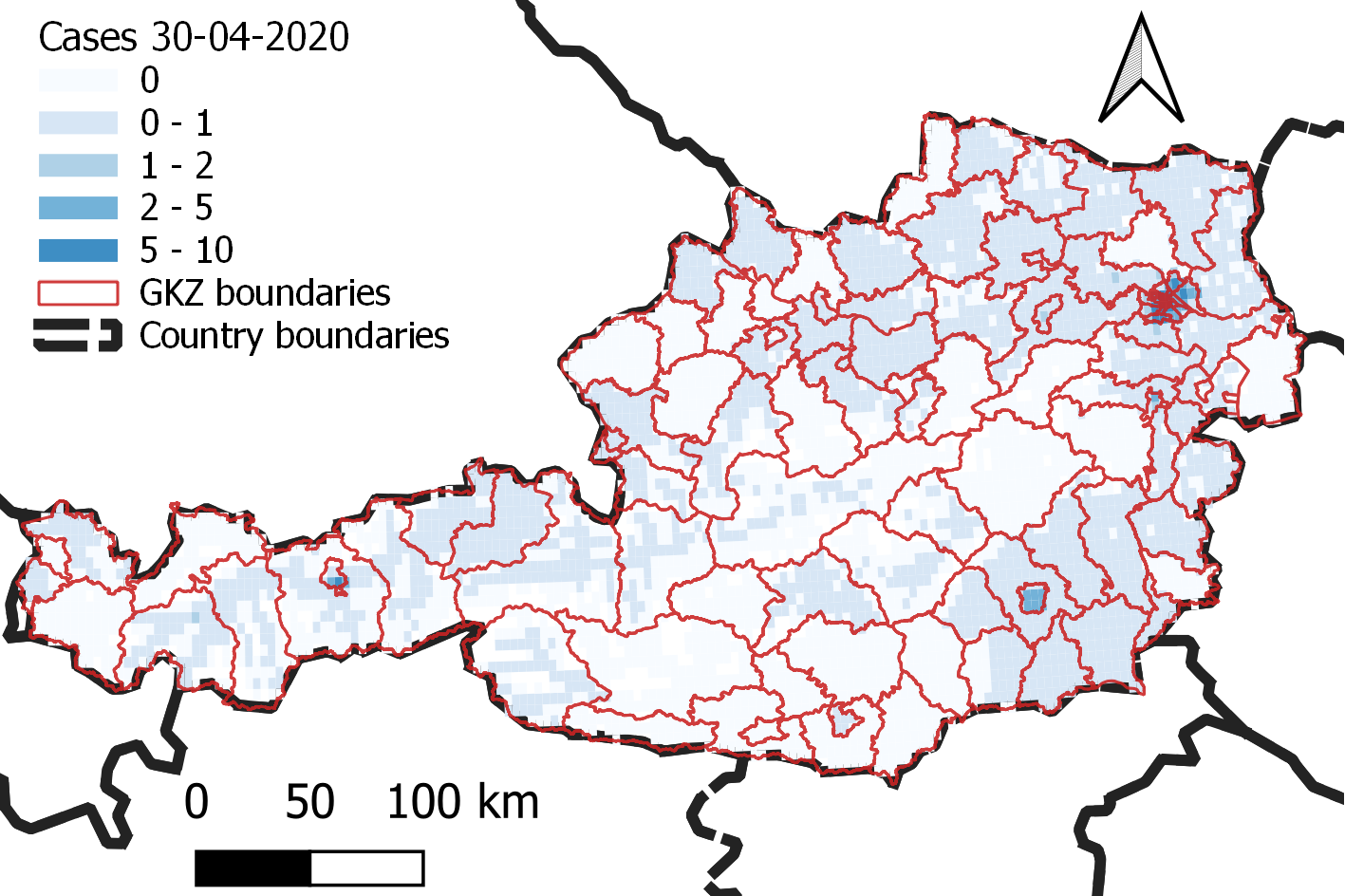}
        
\caption{The number of cases in past 7 days for the dates 15 and 31 March, 15 and 30 April 2020 on the MFAs (left) and on the GKZs (right). It looks like the COVID-19 spread seems to follow more closely the MFAs borders than the GKZ borders.}
    \label{fig:mfaCases}
\end{figure}

As said, the COVID-19 health data for Austria are available only at political district level from the open data repository\footnote{\url{https://github.com/statistikat/coronaDAT}} of the Federal Ministry of social Affairs, Health, Care and Consumer Protection\footnote{\url{https://info.gesundheitsministerium.at/}}.
As it is know that the number of infection cases is proportional to the population size, we redistribute the number of COVID-19 cases on the 4789 zones proportionally to the population size of the zones  within each district.
Figure~\ref{fig:mfaCases} shows the number of cases in the last seven days for four equally spaced dates from 15 March to 30 April 2020. These maps seem to confirm qualitatively the role of the MFAs in the spread of the virus.
To assess this qualitative evidence with a quantitative index, prior to the statistical analysis that will follow, we can use the index of determination (Pearson's correlation ratio) $\eta^2_{Y|G}$ \citep{Fisher25} which measures the dependency of a continuous variable $Y$ in terms of a categorical variable, usually a grouping variable. The index is based on the decomposition of the total variance of $Y$ into the sum of the variance \textit{within} the groups ($\sigma^2_W$) and the variance \textit{between} the groups ($\sigma^2_B$):
$$
\sigma^2 = \sigma^2_W + \sigma^2_B
=\frac{1}{N} \sum_{j=1}^K n_j\sigma^2_j + \frac{1}{N}
\sum_{j=1}^K (\bar x_j - \bar{x}_N)^2 n_k
$$
where $\bar x_j$ and $\sigma^2_j$ are respectively the mean and the variance of group $j$, $\bar{x}_N$ is the global mean, and $N=n_1+n_2+\cdots n_k$, is the total number of observations.
The index of determination is given by the formula
\begin{equation}
    \eta^2_{Y|G} = \frac{\sigma^2_B}{\sigma^2} \quad \in [0,1].
\end{equation}
When the within variance is small it means that the grouping variable creates homogeneous groups in terms of the variability of $Y$, in which case $\eta^2_{Y|G}$ reaches an high value. To compare the effectiveness of the MFAs to capture the dynamics of the pandemic, we calculate the $\eta^2_{Y|G}$ index using the MFAs or GKZs as grouping variables, and we take the sum of number of cases in the past seven days in each MFA or GKZ as response variable $Y$. The results are given in Table~\ref{tab:eta2} and plotted in Figure~\ref{fig:eta2}. The results show that most of the time the value of $\eta^2_{Y|MFA}\geq \eta^2_{Y|GKZ}$ from 15 March 2020 till 20 April 2020, then the reverse is true. This is expected and confirmed by later analysis, because the impact of the MFA can be present only up to few weeks after national lockdown are put in place.

\begin{table}[!htb] 
\centering
\begin{tabular}{rrcc}
  \hline
  MFA & GKZ & date range & calendar week \\ 
  \hline
   0.99 & 0.99 & 2020-03-15 - 2020-03-16 &  11 \\ 
   0.92 & 0.87 & 2020-03-17 - 2020-03-23 &  12 \\ 
   0.98 & 0.97 & 2020-03-24 - 2020-03-30 &  13 \\ 
   0.95 & 0.95 & 2020-03-31 - 2020-04-06 &  14 \\ 
   0.93 & 0.92 & 2020-04-07 - 2020-04-13 &  15 \\ 
   0.84 & 0.84 & 2020-04-14 - 2020-04-20 &  16 \\ 
   0.97 & 0.99 & 2020-04-21 - 2020-04-27 &  17 \\ 
   0.99 & 1.00 & 2020-04-28 - 2020-04-30 &  18 \\ 
   \hline
\end{tabular}
\caption{The index  $\eta^2_{Y|G}$ by calendar week by group $G$  = MFA or $G$ = GKZ. The higher the value of $\eta^2_{Y|G}$, the more $Y$ is determined by $G$.}
\label{tab:eta2}
\end{table}

We now test our hypothesis in two ways. First of all we consider a simple linear model for the log number of cases in the last 7 days and the indicator function $\tt mfaInd$ which takes value equal to 1 if the zone belongs to a MFA and zero otherwise. We also control for the $\tt population$ size of the zone. The model is as follows
\begin{equation}
y_{i,t} = \alpha + \beta \cdot {\tt mfaInd}_i + \gamma \cdot {\tt population}_i,
\label{eq:mod1}
\end{equation}
for $t$ = 15 March 2020, $\ldots$,  30 April 2020, $y_{i,t}$ is the logarithm of the number of COVID-19 cases in the last 7 days for the zone $i$, $i=1, \ldots, 4789$.

Figure~\ref{fig:Coeff} shows the sign of the standardized\footnote{As $\tt mfaInd$ is a categorical variable, to make sense of the standardization we apply Gelman's rule to standardize the $\tt population$ by two standard deviations. See \url{http://www.stat.columbia.edu/~gelman/research/unpublished/standardizing.pdf}.} coefficient of the $\tt mfaInd$ as well as those of $\tt population$ and the constant of the model, along with the corresponding adjusted $R^2$ goodness of fit index for the model. Thanks to the standardization, we can see that, although the population size has the main effect, the belonging to an MFAs is also always significant (the plots shows only coefficients whose $p$-values are less than 0.01).
We use the number of total cases in the past 7 days in order to take into account lagged effects of mobility on the evolution of the pandemic.

It turns out that the MFA indicator function has always a positive coefficient $\beta$ with $p$-value less that 0.001, i.e., always statistically significant. Moreover, the overall fit of the model is acceptable being in the range 0.4-0.5 most of the times during the peak of the epidemic. The fact that towards the end of the period the MFA indicator variable has a coefficient decreasing to zero is due to the fact that the mobility pattern changed due to national lockdown measures (see also Figure~\ref{fig:AustriaMNO}) and thus the MFA are quite different.

Model \eqref{eq:mod1} is capturing an average effect of MFA impact on the number of COVID-19 cases.
To disentangle the impact of each MFA on the pandemic, we use a random effect linear model, also known as mixed effect model \citep{BATES20041} or hierarchical model \citep{Gelman06}. In the form we use it is just a regression with clustered errors that depend on a grouping variable.
In our case, the dependent variable is again $y_{i,t}$ but we replace in \eqref{eq:mod1} the indicator function  $\tt mfaInd$ with the MFA categorical variable to obtain equation \eqref{eq:mod2} below. The MFA variable has 223 different values, so it implies 222 new dummy variables in the model, most of them correlated, and for which there are not much observations per groups as some MFA are made of  two or few more zones. To solve this problem we apply the random effect model as implemented in \citep{JSSv067i01}. In practice, we cluster the observations in groups identified by the political districts ($\tt GKZ$). Exploiting the correlation between the GKZs and the MFAs  it is possible to estimate all the variances of the MFA dummy variables and hence test the significance of each coefficient. The random effect model can be written in this form, where ``$\,  | \,{\tt GKZ}$" means \textit{conditionally to} in the sense of \cite{JSSv067i01}:
\begin{equation}
y_{i,t}    = \alpha + \beta \cdot {\tt mfa}_i + \gamma \cdot {\tt population}_i \,  | \,{\tt GKZ},
\label{eq:mod2}
\end{equation}
\begin{figure}[!htb] 
    \centering
    \includegraphics[width=0.45\textwidth]{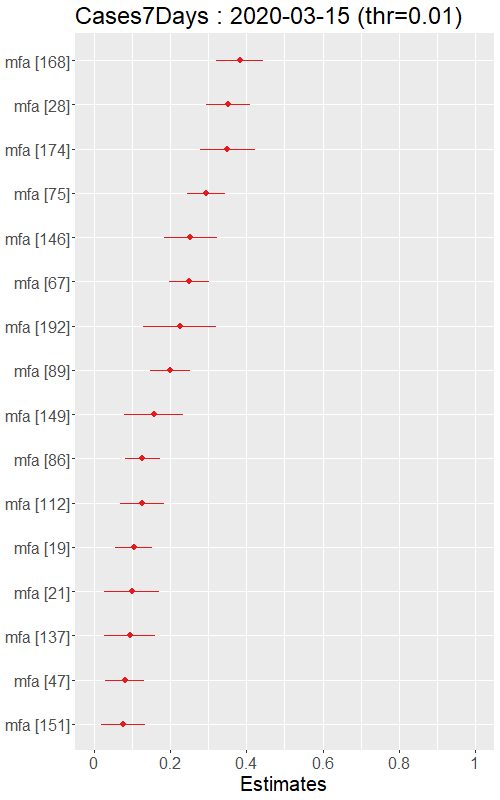}
    \includegraphics[width=0.45\textwidth]{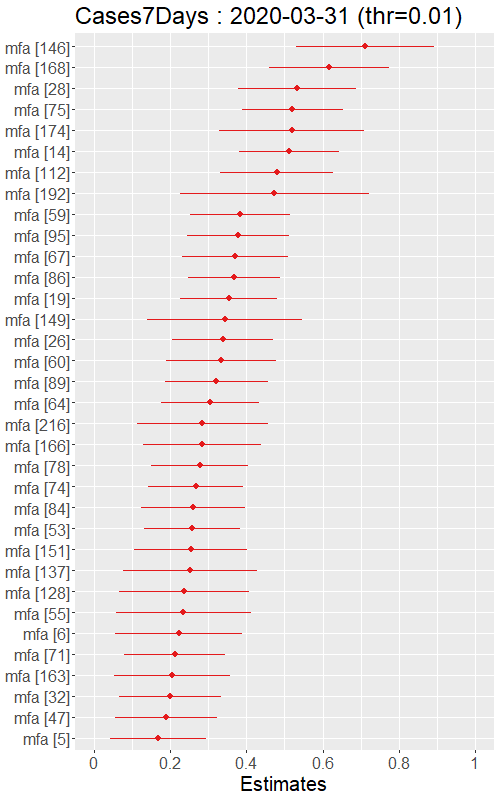}
        \includegraphics[width=0.45\textwidth]{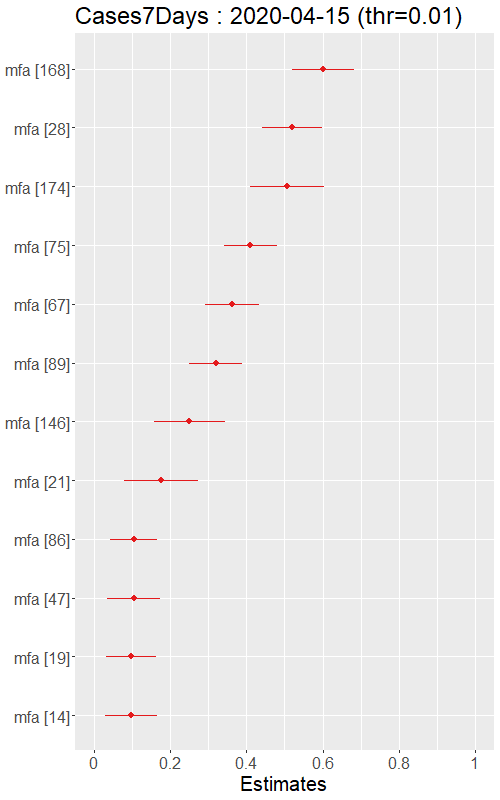}
        \includegraphics[width=0.45\textwidth]{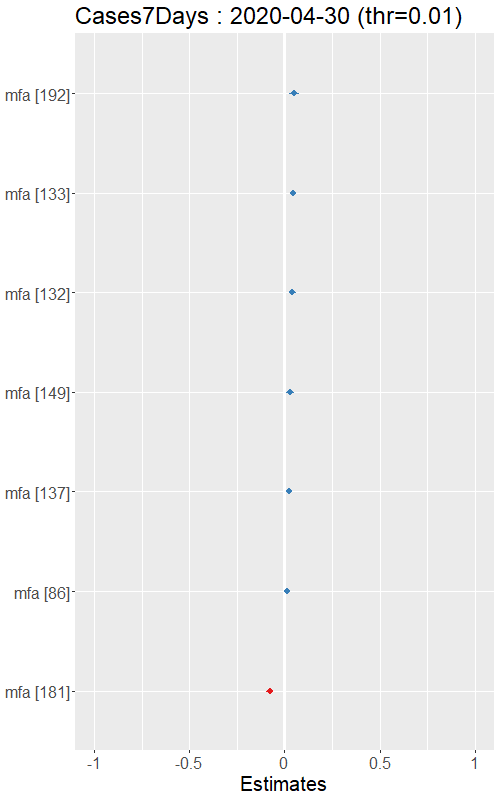}
\caption{Left to right: coefficients of the MFA categorical variable for different dates. Only the coefficients which are statistically different from zero are shown. The only notable MFA is the MFA number 181, for which the sign of the coefficient is negative. The colors of the points are not relevant.}
    \label{fig:randomEff}
\end{figure}
Figure~\ref{fig:randomEff} shows the estimated coefficients, with their respective confidence intervals, for the same four dates of Figure~\ref{fig:mfaCases}. Only the coefficients which are significant at 0.01 are shown. It can be noticed a couple of things: {\it i)} the number of MFAs retained changes through time though those MFAs that remain have a positive coefficient, meaning that being in those particular MFAs increases the number of cases. {\it ii)} it is possible to see that the only MFA which has a negative coefficient if the number 181 but this happens at the end of the period of analysis.  

\begin{figure}[!htb] 
    \centering
    \includegraphics[width=\textwidth]{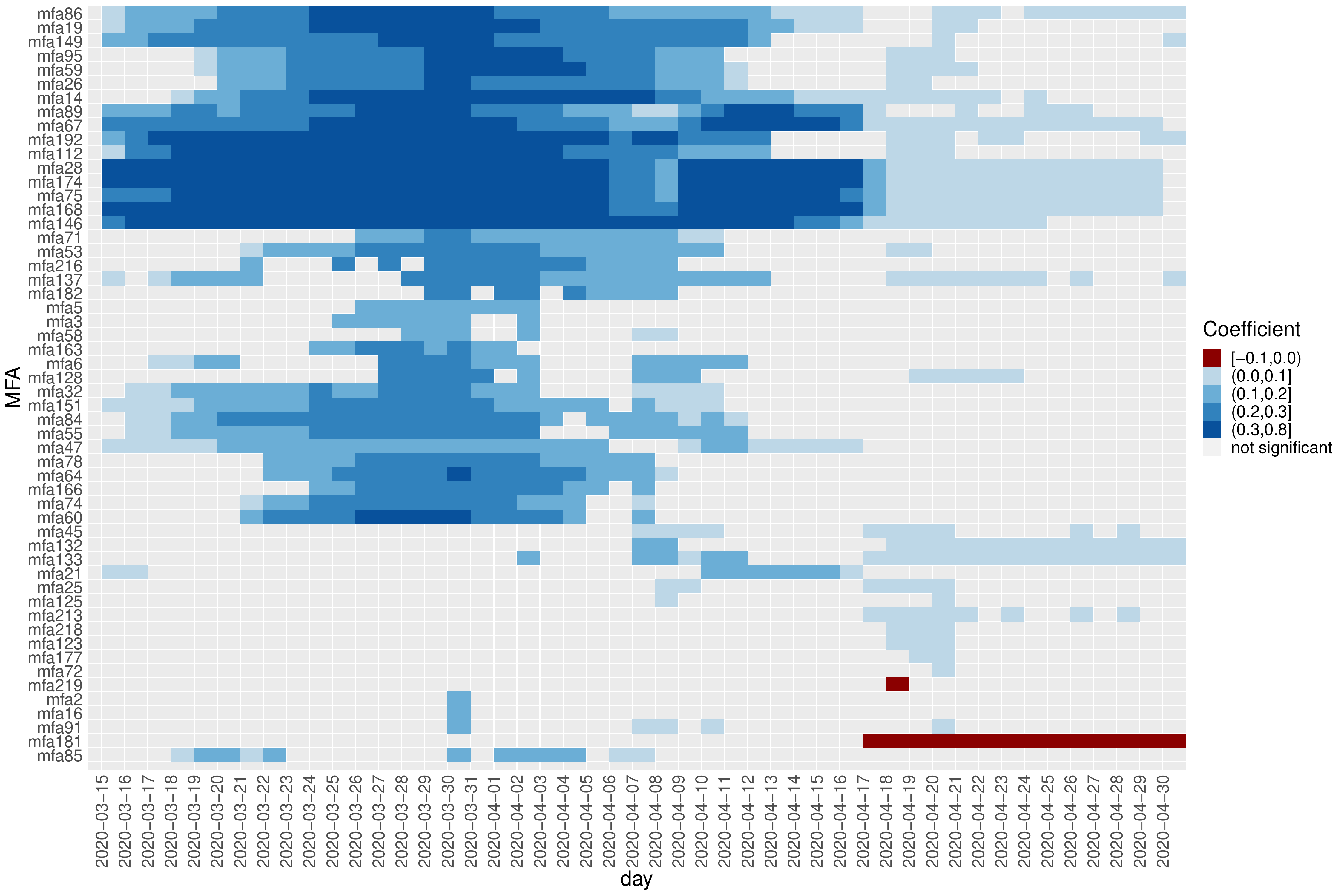}
    \caption{The impact of the MFA on the log number of cases. Blue dots mean that the corresponding MFAs is impacting (increasing) the number of cases; red dots means the contrary. transparent dots means that the corresponding regression coefficient is not significant. }
    \label{fig:CoeffMFA}
\end{figure}

A dynamical view which involves all significant MFAs during the whole period of the analysis is presented in Figure~\ref{fig:CoeffMFA} which shows the heatmap of the corresponding MFAs coefficients which are statistically significant at 0.01 level in all the dates analyzed. It is again clear that the only special MFA is the number 181 that becomes significant and negative starting from 17 April 2020 when all the other MFAs are no longer significant. But around 15 April, the number of cases is also decreasing very quickly towards zero, meaning that there is essentially no signal in the data, i.e., there are no longer cases to analyze.

This MFA number 181 is made of zones in the Wien Stad and Sankt Pölten Land, a small district on the west border to the city of Vienna. The conclusion here seems to be that when the number of cases goes down due to countermeasures (lockdowns, curfews, etc) the impact of the persistent (pre-lockdown) MFAs is clearly vanishing, while until the lockdown, the MFAs are largely driving the pandemic. The negative impact related to the MFA number 181 is mainly due to the lack of cases for all the other MFAs and a few left cases in that MFAs, which naturally leads to a negative sign of the regression coefficient.
 The other MFA which shows a negative coefficient in a single date, is the MFA number 219 that corresponds to the small village of Weiz. As this is one date only, we do not think that this is a statistical evidence so we do not comment any further.

Table~\ref{tab:stats} shows that in terms of population size and therefore number of COVID-19 cases, the MFAs which are not statistically significant are in general smaller. This type of results in quite natural as it means that the areas to be contained or monitored are correctly identified in our approach.
This implies also that the MFA with higher population can be isolated if \textit{i)} they have too many cases, in which case there is high probability of exporting cases; or if \textit{ii)} there are no infections, in which case it is possible to avoid the import of new cases \citep{TR-Connectivity}.

\begin{table}[!htbp] \centering 
\begin{tabular}{@{\extracolsep{5pt}} lrrrr} 
\\[-1.8ex]\hline 
\hline \\[-1.8ex] 
Population & 1st Qu. & Median & Mean & 3rd Qu. \\ 
\hline \\[-1.8ex] 
positive impact MFAs (blue) & $7943$ & $31602$ & $82666$ & $101857$ \\ 
not significant MFAs (transparent) & $1183$ & $3605$ & $24494$ & $15892$\\ 
negative impact MFAs (red) & $6197$ & $12211$ & $12211$ & $18225$\\ \hline \\[-1.8ex] 
Cases & 1st Qu. & Median & Mean & 3rd Qu. \\
positive impact MFAs (blue)  & $441$ & $1792$ & $4508$ &$5230$ \\ 
not significant MFAs (transparent) & $57$ & $167$ & $1118$ &$713$ \\ 
negative impact MFAs (red) & $226$ & $443$ & $443$ & $659$\\ 
\hline \\[-1.8ex] 
\end{tabular} 
\caption{Elementary population size statistics (top) and number of COVID-19 cases (bottom) of the MFAs. The labels correspond to the colors in Figure~\ref{fig:CoeffMFA}.}
\label{tab:stats}
\end{table}

To summarize, this analysis clearly shows the impact of MFAs on COVID-19 spread (along with other causes) and can inform policies in future waves. Indeed, a potential implication of this analysis is that, if new cases arises in a zone belonging to a specific MFA, that MFA should be isolated (or tested extensively) to efficiently contain further spread of the virus, rather than focusing on the whole or single district (when MFAs crosses districts) or instead of blocking the whole country, in the spirit of the zero-covid strategy\footnote{\url{https://www.euronews.com/2021/02/23/what-is-a-zero-covid-strategy-and-could-it-be-implemented-in-europe}.}.

\section{Conclusions}\label{sec:end}
The present work, in line with the literature on functional regions, puts in evidence how administrative borders typically differ from actual commuting patterns, which are shaped by human mobility. An innovative way to reflect human mobility patterns is given by the `\textit{Mobility Functional Areas}' (MFAs). The MFAs are data-driven geographic entities derived from fully anonymised and aggregated data originally provided to the European Commission by several European Mobile Network Operators (MNOs) in the framework of the non-medical interventions to contrast the spreading of COVID-19 pandemic in Europe.\\
Though slightly changing day by day,  MFAs are essentially persistent in time and identify clear intra-weekly patterns.
We have shown how this patterns have changed in Austria following mobility-restrictions imposed by the national authorities to limit the spreading of the virus. Analogous changes in mobility patterns have been measured, with different extents, in all the European countries that we have analysed. In fact, national lockdowns and similar mobility-restriction measures at the local level have not only reduced the volume of mobility but have also drastically changed the mobility patterns and this ``shrinking effect" is clearly reflected in the shape and distribution of the MFAs.\\

The paper shows how, at a very local scale, MFAs can support a more efficient planning of the transport system, as well as its continuous monitoring.
It also introduces several other potential fields of application for the MFAs (i.e. socio-economic, public health, urban planning, environmental risk and pollution, and demography and migration), which should be further investigated in future researches.
Finally, it provides statistical evidences that during the initial phase of the pandemic, the MFAs played an important role in the spread of the virus. On this bases, an enforcement of mobility restrictions based on MFAs rather than administrative areas would have been more effective in limiting both the spreading of the virus and the socio-economic impact of the restrictions. That said, the authors acknowledge that from an operative point of view, the enforcement of mobility restrictions based on administrative borders (which are well recognised by citizens and reflect administrative and law-enforcement areas) would be much easier to implement than one based on MFAs, which very often cross these administrative borders. For example, in case of a local outbreak in a municipality, instead of applying restrictive measures in a whole region or country health authorities can only apply them in the municipalities that belong to the same MFA. For this reason, coordination efforts between local and regional authorities are required.\\

While this study focuses on Austria, the full study comprises the analysis of Sections~\ref{sec:MFA} to~\ref{sec:Vienna}, for 15 European countries as shown in Table~\ref{tab:MFAdata} and the definition of the MFAs and the corresponding shapefiles are available through the KCMD Dynamic Data Hub\footnote{\url{https://bluehub.jrc.ec.europa.eu/catalogues/info/dataset/ds00170}} for further analysis by scholars.

\begin{acknowledgements}
The authors acknowledge the support of European MNOs (among which A1 Telekom Austria Group, Altice Portugal, Deutsche Telekom, Orange, Proximus, TIM Telecom Italia, Telefonica, Telenor, Telia Company and Vodafone) in providing access to aggregate and anonymised data  as an invaluable contribution to the initiative.\\
The authors would also like to acknowledge the GSMA\footnote{GSMA is the GSM Association of Mobile Network Operators.}, colleagues from DG CONNECT\footnote{DG Connect: The Directorate-General for Communications Networks, Content and Technology is the European Commission department responsible to develop a digital single market to generate smart, sustainable and inclusive growth in Europe.} for their support and colleagues from Eurostat\footnote{Eurostat is the Statistical Office of the European Union.} and ECDC\footnote{ECDC: European Centre for Disease Prevention and Control. An agency of the European Union.}  for their input in drafting the data request. \\
Finally, the authors would also like to acknowledge the support from JRC colleagues, and in particular the E3 Unit, for setting up a secure environment, a dedicated \textit{Secure Platform for Epidemiological Analysis and Research} (SPEAR) enabling the transfer, host and process of the data provided by the MNOs; as well as the E6 Unit (the Dynamic Data Hub team) for their valuable support in setting up the data lake.
\end{acknowledgements}


\section*{Declarations}

\paragraph{Funding:} Not applicable.

\paragraph{Availability of data and material:} Original mobile network data are not available and must be agreed individually with each MNO.
The MFA's definitions and the corresponding shapefiles and mapping files are available at this link \url{https://bluehub.jrc.ec.europa.eu/catalogues/info/dataset/ds00170}.

\paragraph{Code availability:} All code written in R language \citep{RCore}, available on request to the authors.

\paragraph{Conflict of interest:} No potential competing interest is reported by the authors.

\paragraph{Authors' contribution:}\phantom{.}\\
SM Iacus: Manuscript writing, Content planning, Literature review, Modelling and Data Analysis.\\
M. Vespe: Content planning and Manuscript editing.\\
S. Spyratos: Literature review, Manuscript editing, Maps building.\\
C. Santamaria, F. Sermi and D. Tarchi: Manuscript editing.\\

\section*{Appendix}

\begin{table}[!htb] 
    \centering
\scriptsize{    
\begin{tabular}{lc|lc|lc| c }
        Country & (ISO2) & highest granularity & used   &NUTS 3 & used   & date range  \\
\hline
        Austria & (AT) & grid 16-18 km$^2$ &  4789   & districts & 35    & 01/02/20 - 29/06/20\\
        Belgium & (BE) & postal code areas & 1131 & regions & 44 & 11/02/20 - 29/06/20\\
        Bulgaria & (BG) & grid 16-18 km$^2$ & 4615  & provinces & 28    &  01/02/20 - 27/06/20\\
        Czechia & (CZ) & regular grid & 4014 & regions & 14 & 01/01/20 - 28/06/20\\
        Denmark & (DK) & municipalities& 98 & provinces & 12  & 02/02/20 - 07/06/20\\
        Estonia & (EE) &municipalities & 79 & counties & 5  & 14/02/20 - 07/06/20\\
        Spain & (ES)& municipalities &  6893  & provinces & 59 &   01/02/20 - 29/06/20 \\ 
        Finland & (FI) &municipalities & 310 & provinces & 19  & 02/02/20 - 07/06/20\\
        France & (FR)& municipalities& 1426 & departments & 96 &01/01/20 - 23/06/20\\
        Greece & (GR) &grid 25 km$^2$  & 6240 & prefectures & 53 & 15/05/20 - 30/06/20\\
        Croatia & (HR) &grid 17 km$^2$ &1384 & counties & 22 & 01/02/20 - 19/06/20\\
        Italy & (IT) &census areas & 8051 & provinces & 110  & 01/01/20 - 29/06/20\\
        Norway & (NO) &municipalities & 422 & counties & 18 & 02/02/20 - 07/06/20\\
        Norway & (NO) & municipalities & 356 &  counties & 18 & 20/01/20 - 21/06/20\\
        Sweden & (SE) &municipalities &  290 & counties & 21   & 02/02/20 - 07/06/20\\
        Slovenia & (SI) & grid 16-18 km$^2$ & 1248  & provinces & 12 &  01/02/20 - 29/06/20\\
    \end{tabular}}
    \caption{Data used in the analysis of the MFAs. Some areas (like overseas territories) are excluded.}
    \label{tab:MFAdata}
\end{table}
\bibliographystyle{spbasic}      % basic style, author-year citations
\bibliography{refbib}   % name your BibTeX data base

%\bibliography{refbib}
%\bibliographystyle{chicago}

\end{document}